\documentclass[
aps,
prb,
twocolumn,
superscriptaddress,%
showpacs,%
preprintnumbers,%
floatfix%
]{revtex4-2}

\usepackage{amsmath,amsfonts,amssymb}
\usepackage{graphicx}
\usepackage{graphpap}

\usepackage[colorlinks=true, allcolors=blue]{hyperref}
\usepackage{todonotes}
\usepackage{ulem}
\usepackage{nicefrac}

\usepackage{xcolor}

\usetikzlibrary{arrows.meta}
\usetikzlibrary{shapes.geometric}
\tikzset{%
  >={Latex[width=2mm,length=2mm]},
            base/.style = {rectangle, rounded corners, draw=black,
                           minimum width=4cm, minimum height=1cm,
                           text centered, font=\sffamily},
  initialization/.style = {base, fill=blue!30},
     quantum_run/.style = {base, fill=red!30},
     classic_run/.style = {base, fill=green!30},
         process/.style = {base, minimum width=2.5cm, fill=orange!15,
                           font=\ttfamily},
}
\usepackage{varwidth}

\begin{document} 

\title{Variational quantum-algorithm based self-consistent calculations for the two-site DMFT model on noisy quantum computing hardware}

\author{Jannis Ehrlich}
\email{jannis.ehrlich@iwm.fraunhofer.de}
\affiliation{Fraunhofer-Institut für Werkstoffmechanik IWM, Wöhlerstraße 11, Freiburg, Germany}
\author{Daniel F. Urban}
\affiliation{Fraunhofer-Institut für Werkstoffmechanik IWM, Wöhlerstraße 11, Freiburg, Germany}
\affiliation{Freiburger Materialforschungszentrum, Universität Freiburg, Stefan-Meier-Straße 21, Freiburg, Germany}
\author{Christian Elsässer}
\affiliation{Fraunhofer-Institut für Werkstoffmechanik IWM, Wöhlerstraße 11, Freiburg, Germany}
\affiliation{Freiburger Materialforschungszentrum, Universität Freiburg, Stefan-Meier-Straße 21, Freiburg, Germany}

\begin{abstract}
Dynamical Mean Field Theory (DMFT) is one of the powerful computational approaches to study electron correlation effects in solid-state materials and molecules. Its practical applicability is, however, 
limited by the quantity of numerical resources required for the solution of the underlying auxiliary Anderson impurity model. Here, the one-to-one mapping between electronic orbitals and the state of a qubit register suggests a significant computational advantage for the use of a Quantum Computer (QC) for solving this task. In this work we present a QC approach to solve a two-site DMFT model based on the Variational Quantum Eigensolver (VQE) algorithm. We analyse the propagation of stochastic and device errors through the algorithm and their effects on the calculated self-energy. Therefore, we systematically compare results obtained on simulators with calculations on the IBMQ Ehningen QC hardware. We suggest a means to overcome unphysical features in the self-energy which already result from purely stochastic noise. Based hereon, we demonstrate the feasibility to obtain self-consistent results of the two-site DMFT model based on VQE simulations with a finite number of shots. 
\end{abstract}

\maketitle

%
%

\section{Introduction}
The understanding and design of functional materials and molecules is nowadays widely supported by computational studies. On the atomic scale, density functional theory (DFT) is the widely used tool to calculate various physical properties for a large class of materials at reasonable accuracy and numerical cost \cite{Martin2020}. However, DFT cannot correctly capture the physics of strongly correlated electron systems. This shortcoming originates from the exchange-correlation functional, for which only approximate forms are available, such that the interactions between  electrons are only treated on a mean field level.

One approach to overcome these limitations is the dynamical mean field theory (DMFT) \cite{Metzner1989,Georges1992}. It  exploits that electronic interactions are strongest on short distances as for example in the partially filled \(d\)-shells of transition metal atoms or \(f\)-shells of rare earth atoms.
DMFT is based on the equivalence of two notions of a correlated orbital: (i) as an impurity connected to an uncorrelated bath in terms of an Anderson impurity model (AIM) \cite{Anderson1961} and (ii) as part of a regular crystal lattice. These two notions lead to a self-consistency condition formulated in terms of Green's functions (GF). 

In practice, DMFT calculations are routinely performed for real materials  with finite system sizes \cite{Georges1996,Held2007,Kotliar2004,Kotliar2006,Nilsson2017,Watzenbock2022, Chen2022, Karolak2010} and provide a good insight into the physics of strong electron correlations. The bottleneck in these calculations is the solution of the AIM 
for which various methods have been developed and extensively studied. The solution by exact diagonalization (ED) is limited to \(\sim 24\) orbitals \cite{Sangiovanni2006} due to the exponential scaling of the Hamiltonian in the number of orbitals, while some modifications even allow the treatment of up to \(\sim 100\) orbitals \cite{Lupo2021}. Quantum Monte Carlo (QMC) methods work best for relatively large temperatures, such that the interesting low-temperature regime of competing two-particle effects becomes challenging. The Density Matrix Renormalization Group (DMRG) \cite{Garcia2004}, however, is limited to a small bond dimension and thus a small number of interacting orbitals. There exist further approaches to solve the AIM, e.g. the numerical renormalization group \cite{Zitko2009} method, which come with their own limitations.

In addition to the above mentioned approaches which are applicable on classical computers, first hybrid classical-quantum computing algorithms were presented recently which have the potential to solve the AIM \cite{Bauer2016, Jaderberg2020, Keen2020, Kreula2016, Rungger2019, Endo2020}. In the spirit of Feynman's idea \cite{Feynman1982}, these algorithms use quantum computers to simulate equivalent quantum systems. In these approaches, each fermionic spin-orbital of the original system is mapped to one qubit, such that the exponential scale of the Hilbert-space becomes a linear scale in the number of qubits. Current QC devices already have more than one hundred qubits, and thus, in principle, they allow the simulation of more than one hundred orbitals, going beyond the capability of any classical computer. However, even the lowest noise rates that were achieved on a real device so far are still on such a scale, that significant additional resources for error mitigation are required, to obtain reasonable results when tens of qubits are used.

The quantum-classical hybrid approaches to obtain the GF as a solution of a model Hamiltonian like the AIM can be categorized mainly into two different types. First, in \textit{time evolution} approaches \cite{Bauer2016, Jaderberg2020, Keen2020, Kreula2016, Endo2020, Libbi2022, Steckmann2023, Gomes2023} the real-time GF is calculated by simulating directly the time evolution of the system generated by its Hamiltonian. The spectral function can then be obtained via a Fourier transformation. The second type of approaches are based on the \textit{Lehmann representation} of the GF and require the computation of probabilities for transitions from the ground state to excited states as well as the energy differences between those states. Here, one can use \textit{variational optimization} to obtain energies and quantum circuits for the ground and excited states \cite{Rungger2019, Endo2020, Jamet2021}, which are then used to calculate the transition rates on a quantum computer. Alternatively, \textit{subspace expansion} methods \cite{Jamet2022, Rizzo2022} start from an (approximate) ground state, measure the matrix elements of its expansion in a quantum subspace, classically diagonalize the generalized eigenvalue problem and use this classical result for the construction of the Lehmann GF.

The minimal realization of a DMFT approach is the two-site DMFT model which was introduced in Ref.\ \cite{Potthoff2001}. This model was already considered as a test case for hybrid classical-quantum computer algorithms, namely the time-evolution approaches in Refs.\ \cite{Kreula2016, Jaderberg2020, Keen2020, Steckmann2023} and the Lehmann representation approach in Ref.\ \cite{Rungger2019}. In the case of time-evolution, the required long circuits include high two-qubit gate counts and lead to a significant noise rate on current hardware. Thus, in Refs.~\cite{Kreula2016, Jaderberg2020, Keen2020} the analytically known form of the time-dependent GF for the half-filled two-site model was fitted to the measured GF, such that the Fourier transformation could be performed by inserting the fitted parameters into the frequency dependent GF. Steckmann et al.\ \cite{Steckmann2023} used a generic discrete Fourier transformation on the real-time GF and thus do not rely on the known analytic form. However, in the derivation of their simplified quasi-particle weight calculation, which directly enforces a constraint to the self-energy in order to correct unphysical features, they still exploit the specific analytic form and sum rule of the particle-hole symmetric case. Moreover, the classical Cartan-decomposition scheme they used for the Hamiltonian has an exponential scaling with the system size. Therefore, this approach cannot easily be generalized to models of many more than two sites. \\

In this work we explore some encountered practical issues in the use of the Lehmann-based hybrid classical-quantum approach to DMFT on Noisy Intermediate-Scale Quantum (NISQ) computer systems by studying the two-site DMFT model \cite{Potthoff2001} as an example.
The paper is organized as follows. We first introduce the general DMFT approach, as well as the two-site DMFT specifics in Sec.\ \ref{Sec:DMFT}. In Sec.\ \ref{Sec:QC} we then discuss the mapping of the AIM to a quantum computer and our approach to obtain its full GF. Section\ \ref{Sec:NoiseModels} presents the different noise models and the error mitigation strategy used in our study. In Sec.\ \ref{Sec:Results} we discuss the results for the different steps of the quantum algorithm with increasing complexity in the QC contribution, including calculations on cloud quantum computers. 
First, we investigate only the eigenvalues and eigenstates (Sec.\ \ref{Sec:Energies}), then we construct the GF and the self-energy (Sec.\ \ref{Sec:GF}). Since unphysical features appear in the self-energy, we introduce an efficient fitting procedure to avoid artifacts arising from the use of approximate results. We estimate the required quantum resources for a full self-consistent DMFT simulation (Sec.\ \ref{Sec:Scaling}), then perform this simulation (Sec.\ \ref{Sec:Self_cons}), and apply the developed work flow to investigate the Mott-insulator transition of the two-site DMFT model (Sec.\ \ref{Sec:Mott_MIT}).
Throughout our analysis we verify the correctness of the hybrid algorithm by an ideal noiseless simulation based on linear algebra. In a second step, we account for the probabilistic nature of quantum mechanics which manifests itself in the measurement process on the quantum computer and study the effect of shot noise on the performance of the algorithm. The influences of gate, readout and decoherence errors are subsequently studied using a simulator with tunable error rates. Our analysis is complemented by results obtained on the IBMQ System One Ehningen, a superconducting 27-qubit Falcon r5.11 processor.

%
%

\section{Dynamical Mean Field Theory}\label{Sec:DMFT}

\begin{figure*}
\includegraphics[width=2\columnwidth]{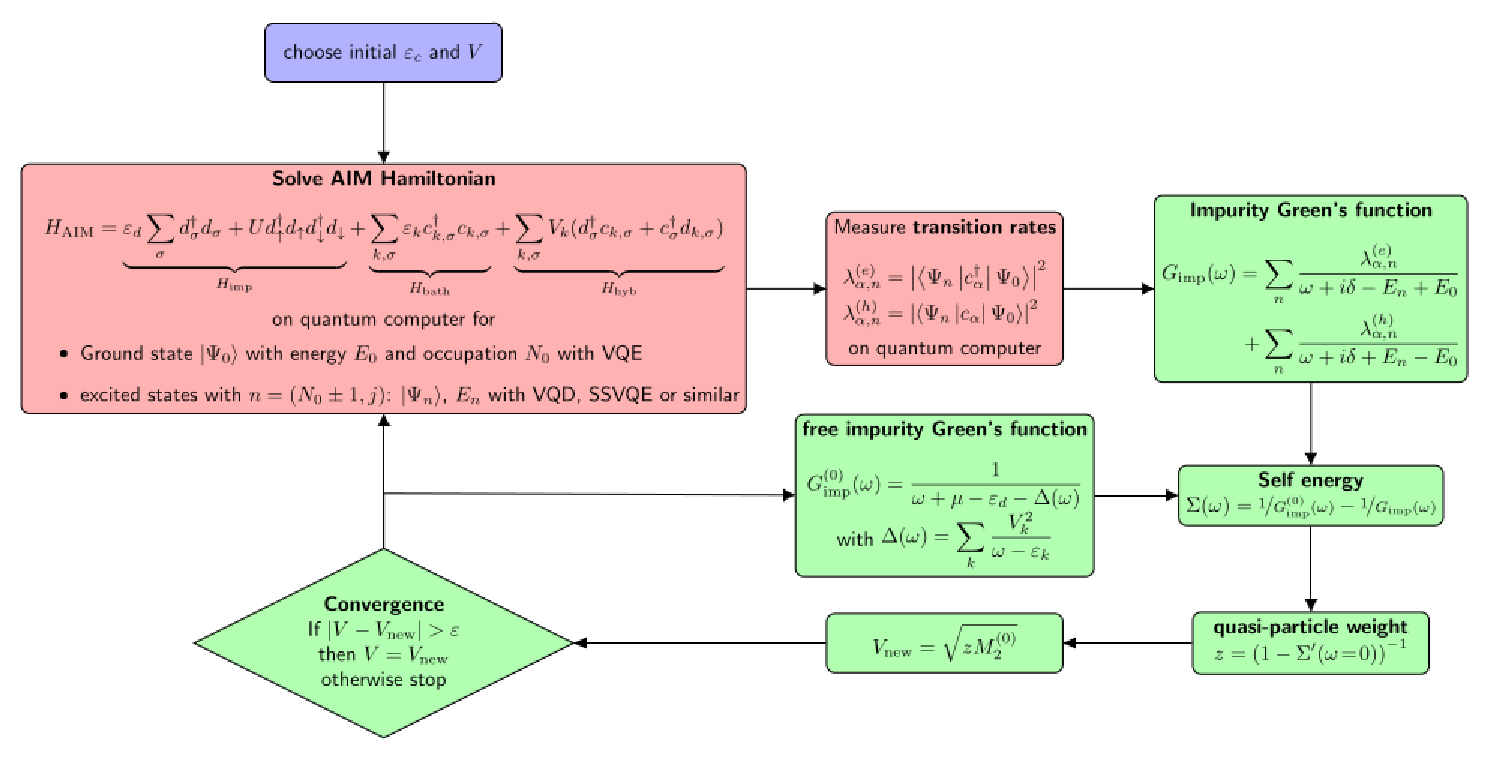}
\caption{Schematic illustration of the inner self-consistency loop for the two-site DMFT model for determining the hybridization strength \(V\) while keeping the impurity occupation \(n_\text{imp}\) constant. Red boxes indicate operations performed on a quantum computer while green boxes indicate purely classical computations.}
\label{Fig:DMFT_loop}
\end{figure*}

The Hubbard model provides the simplest model for the description of a correlated electron system \cite{Hubbard1963}. Its Hamiltonian
\begin{align}
H_{\text{Hub}} = -t \sum_{\langle i,j \rangle, \sigma} c_{i,\sigma}^\dagger c_{j,\sigma} + U \sum_{i} n_{i, \uparrow} n_{i, \downarrow}
\end{align}
includes the movement (hopping) of electrons between neighboring lattice sites $i$ and $j$ 
with hopping energy \(t\) and the interaction between electrons with strength \(U\), which acts only between two electrons with opposite spin when they occupy the same lattice site. Here \(c_{i,\sigma}^\dagger\) and \(c_{i,\sigma}\) are the creation and annihilation operators of an electron with spin \(\sigma\) at site \(i\), respectively, and \(n_{i,\sigma}=c_{i,\sigma}^\dagger c_{i,\sigma}\) is the particle number operator on site \(i\). In spite of its seeming simplicity, this model is not analytically solvable in more than one spatial dimension.

The DMFT divides the solution of the Hubbard model into two tasks: First, the full lattice is described by GF \(G_\text{lat}\) which requires self-energies \(\Sigma\) which are assumed to be local. Second, the self-energy can be efficiently calculated by an auxiliary model consisting of a single site with interacting orbitals which is coupled to an effective bath. This bath has to be parametrized in such a way, that it represents the crystal lattice surrounding the specific site. Coupling the two tasks has the additional challenge, that the solution has to be determined self-consistently. (For reviews on DMFT see e.g.\ Ref.\ \cite{Georges1996, Kotliar2006}.) The auxiliary model system is equivalent to the AIM with the Hamiltonian
\begin{align}\label{eq:AIM}
H_{\text{AIM}} &= H_{\text{imp}} + H_{\text{bath}} + H_{\text{hyb}}\\
H_{\text{imp}} &= \varepsilon_d \sum_{\sigma} d_{\sigma}^\dagger d_{\sigma} + U d_{\uparrow}^\dagger d_{\uparrow}d_{\downarrow}^\dagger d_{\downarrow}\\
H_{\text{bath}} &= \sum_{k, \sigma} \varepsilon_k c_{k, \sigma}^\dagger c_{k, \sigma}\\
H_{\text{hyb}} &= \sum_{k,\sigma} V_k (d_{\sigma}^\dagger c_{k,\sigma} + c_{k, \sigma}^\dagger d_{\sigma}).
\end{align}
The AIM describes a single impurity site 
\(d^\dagger_\sigma\left|0\right>\)
with single-particle energy \(\varepsilon_d\) where electrons of opposite spin interact with strength \(U\). This impurity is connected to a bath of non-interacting sites 
\(c^\dagger_{k,\sigma}\left|0\right>\) with energy \(\varepsilon_k\) via a hybridization strength \(V_k\). 

The central quantity in DMFT is the GF which has to be determined self-consistently. Based on the equivalence of the two models, the lattice GF on the impurity site \(G_{\text{lat}}\) has to be identical to the impurity GF \(G_\text{imp}\) in the AIM. This directly implies the equivalence of the self-energy of both systems. 

The simplest possible DMFT model consists of only one impurity and one bath site, \(H_{\text{bath}} = \sum_{\sigma} \varepsilon_c c_{\sigma}^\dagger c_{\sigma}\). This so-called two-site DMFT, which we consider in this paper, was studied in detail by Potthoff \cite{Potthoff2001}. 
Here, we review the essential ingredients of the corresponding algorithm. Two specific conditions for self-consistency,
\begin{align}\label{Eq:sc_cond}
n_\text{imp} = n \quad \text{and} \quad V^2 = zM^{(0)}_2,
\end{align}
were derived in \cite{Potthoff2001}. These conditions demand that the occupation of the correlated site has to be equal in both formulations of the problem and the hybridization strength \(V\) must be equal to the product of the quasi-particle weight \(z\) and the variance of the non-interacting density of states, i.e.\ the bandwidth of the bath electrons, \({M^{(0)}_2 =\sum_{i\neq j}t_{ij}^2 = \int d\omega \omega^2 \rho_0(\omega)}\).

A practical calculation proceeds according to the following protocol:
\begin{itemize}
	\item[(0)] Start with an initial guess for the bath energies \(\varepsilon_c\) and the hybridization strength \(V\).
	\item[(1)] Solve the AIM with parameters \(\varepsilon_c\) and \(V\) under the constraint $V^2 = zM^{(0)}_2$.
	\item[(2)] Calculate the occupation of the impurity given by \(n_\text{imp} = \frac{2}{\pi}\int_{-\infty}^0 \text{d}\omega\, \text{Im}\,G_\text{imp}(\omega + i0^+)\), where \(0^+\) refers to an infinitesimal shift to the positive half plane.
	\item[(3)] Calculate the band-filling (cf. Ref.\ \cite{Potthoff2001}, Eq.\ (25)) via \({n=2 \int_{-\infty}^0 \text{d}\omega \rho_0[\omega + \mu - \Sigma(\omega)]}\),
	\item[(4)] Determine a new \(\varepsilon_c\) (using a classical optimizer) that reduces \(\delta n=\left|n_\text{imp}-n\right|\) and restart from step 1 until $\delta n$ is sufficiently small. 
\end{itemize} 

Besides the outer self-consistency loop (1)--(4) for the occupation number, an inner self-consistency cycle is needed in step (1) in order to ensure Potthoff's second condition $V^2 = zM^{(0)}_2$. Here, the protocol which is illustrated schematically in Fig.\ \ref{Fig:DMFT_loop} consists of the following steps 
\begin{itemize}
	\item[(1.1)] Solve the AIM (Eq.\ \eqref{eq:AIM}) for the chosen parameters \(\varepsilon_c\) and  
	\(V\) and obtain eigenenergies \(E_n\) and eigenstates \(\left|\Psi_n\right>\) of $H_{\rm AIM}$. 
	\item[(1.2)] Calculate the impurity GF based on the Lehmann representation
	\begin{equation}
	G^\sigma_\text{imp}(\omega)= \sum_n \frac{\left|\left< \Psi_n \left| d^\dagger_\sigma \right| \Psi_0\right> \right|^2}{\omega + i\delta - E_n + E_0} \\ 
	+ \sum_n \frac{\left|\left< \Psi_n \left| d_\sigma \right| \Psi_0\right> \right|^2}{\omega + i\delta + E_n - E_0} \label{Eq:ImpLehmann}
	\end{equation}
	\item[(1.3)] Calculate the free impurity GF by
	\begin{align}
	 G_\text{imp}^{(0)}(\omega) = \frac{1}{\omega + \mu -\varepsilon_d - \Delta(\omega)}
	\end{align}
	with hybridization function \(\Delta(\omega) = \sum_k \frac{V_k^2}{\omega - \varepsilon_k}\)
	\item[(1.4)] Calculate the self-energy by the inverted Dyson equation
	\begin{align}\label{Eq:Dyson}
	\Sigma(\omega) = [G^{(0)}_\text{imp}(\omega)]^{-1}-[G_\text{imp}(\omega)]^{-1}
	\end{align}
	\item[(1.5)] Obtain the quasi-particle weight
	\begin{align}\label{Eq:QP_weight}
	z = \left({1 - \left. \frac{\text{d}\Sigma(\omega)}{\text{d}\omega} \right|_{\omega=0}}\right)^{-1}
	\end{align}
	\item[(1.6)] Update the hybridization strength \(V = \sqrt{zM^{(0)}_2 }\)
	\item[(1.7)] Start from step 1.1 until a self-consistent \(V\) is obtained
\end{itemize}
For a detailed discussion including the derivation of equations see Refs.\ \cite{Potthoff2001,Rungger2019}. In practice, it is convenient to stabilize the convergence by using an appropriate mixing when updating the hybridization strength in step (1.6)  before starting the next iteration cycle. In our case of fluctuating values, we take a weighted mean of the last four measured values, where the weight corresponds to the relative difference with respect to the unweighted mean value.

In our simulations we construct the GF for the spin-up operators, which is identical to the spin-down GF according to the spin-symmetry of the system. In the Lehmann representation, Eq.\ \eqref{Eq:ImpLehmann}, only the states with one electron added (removed) relative to the ground state (GS) contribute to the first (second) term. Thus, in the following we refer to the states with one electron more (less) relative to the GS by electron (hole) states and denote the terms of the GF correspondingly.

In the special case of half-filling \(n_\text{imp}=n=1\), the condition for the occupation is always fulfilled, such that only the inner optimization loop for \(V\) has to be performed. In this case, \(\mu=\varepsilon_c=\nicefrac{U}{2}\), the electron (hole) states are those with \(N=3\) (\(N=1\)) electrons in total, and the problem can be solved analytically \cite{Potthoff2001}. The exact solution for the self-energy at half-filling reads
\begin{align}
\Sigma(\omega)= \frac{U}{2}+\frac{U^2}{8}\left(\frac{1}{\omega - 3V} + \frac{1}{\omega + 3V} \right)
\end{align}
and the hybridization strength is given by
\begin{align}\label{Eq:V_halffilled}
V = \begin{cases}
	\sqrt{M_2^{(0)}}-\frac{U^2}{36} \quad \text{ for } U< U_c\\
	0 \quad \text{ else, } 
	\end{cases}
\end{align}
with the critical interaction parameter \(U_c := 6\sqrt{M^{(0)}_2} \) indicating a Mott-insulator transition. For \(U<U_c\) the system is metallic with two singularities of the self-energy at \(\omega = \pm 3V\). When $U$ approaches $U_c$ from below, the two singularities move closer towards \(\omega=0\), until they meet at \(U=U_c\) so that the system becomes insulating due to the strong interactions between the electrons for \(U>U_c\).
The analytic solution for the quasi-particle weight at half-filling is given by
\begin{align}
z= \begin{cases}
	1-\left(\frac{U}{U_c}\right)^2 \quad \text{ for } U<U_c\\
	0 \quad \text{ else.}
	\end{cases}
\end{align}

%
%

\section{Quantum Computing Approach}\label{Sec:QC}

\begin{figure}
\vspace{10pt}
\includegraphics[width = 0.95\columnwidth]{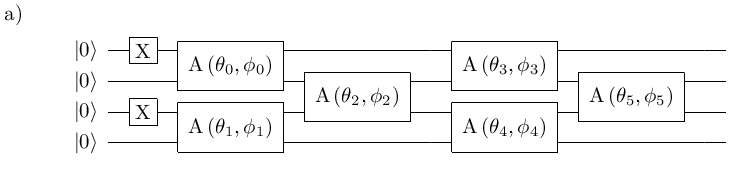}
\includegraphics[width = .45\columnwidth]{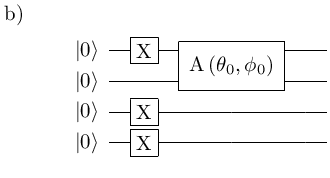}
\includegraphics[width = .45\columnwidth]{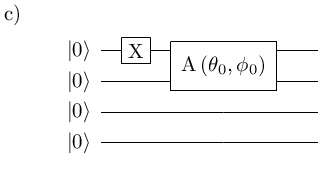}
\includegraphics[width = \columnwidth]{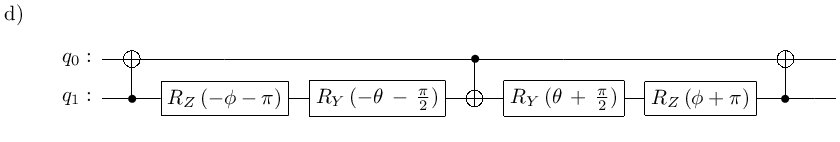}
\caption{Quantum circuits used for the computation of (a) the ground state, (b) electron states with spin up, and (c) hole states with spin up. 
(d) Decomposition of the A-gate into single qubit rotations and CNOT gates.
}\label{Fig:Quantumcircuits}
\end{figure}

We want to perform the most demanding task of DMFT calculations on a QC, namely
the determination of the impurity GF \(G_{\text{imp}}\). We use the Jordan Wigner transformation \cite{Jordan1928} for the mapping between the Fermionic creation/annihilation operators and the Pauli spin operators 
\begin{equation}
   c_i^{\dagger} = \frac{1}{2} \left(\prod_{j=1}^{i-1}\sigma_j^z\right)\left(\sigma_i^x - i \sigma_i^y\right) 
\end{equation} 
and
\begin{equation}
   c_i = \frac{1}{2} \left(\prod_{j=1}^{i-1}\sigma_j^z\right)\left(\sigma_i^x + i \sigma_i^y\right),
\end{equation} 
with \(\sigma_i^x\), \(\sigma_i^y\), \(\sigma_i^z\) denoting the \(x,\ y,\ z\) Pauli operators, respectively. By this transformation we assign each orbital of the model to one qubit in the following order: \(\left(d_\uparrow, c_\uparrow, d_\downarrow, c_\downarrow\right) \rightarrow \left(q_0, q_1, q_2, q_3\right)\). 
With this mapping, the AIM Hamiltonian (Eq.\ \eqref{eq:AIM}) is written in terms of Pauli operators as
\begin{align}
H&=&\mathbf{I}\left(\frac{U}{4}+\varepsilon_d+\varepsilon_c-2\mu\right)
-\left(\sigma_1^z+\sigma_3^z\right)\left(\frac{\varepsilon_c}{2}-\frac{\mu}{2}\right)\nonumber\\
&&
-\left(\sigma_0^z + \sigma_2^z\right)\left(\frac{U}{4} + \frac{\varepsilon_c}{2}-\frac{\mu}{2}\right) \nonumber\\
&&
+ \frac{V}{2}\left(\sigma_0^x\sigma_1^x+\sigma_2^x\sigma_3^x+\sigma_0^y\sigma_1^y+\sigma_2^y\sigma_3^y\right)
\end{align}
As mentioned above, we will consider the special case of half-filling for which the parameters
\(\varepsilon_d=0\) and \(\varepsilon_c=\nicefrac{U}{2}=\mu\) \cite{Potthoff2001}. 
The Hamiltonian thus simplifies to
\begin{equation}
H=\frac{V}{2}\left(
\sigma_0^x\sigma_1^x+\sigma_2^x\sigma_3^x
+
\sigma_0^y\sigma_1^y+\sigma_2^y\sigma_3^y\right)
+\frac{U}{4}\sigma_0^z\sigma_2^z.
\end{equation}
Note that in this special case of half-filling V is determined via Eq.\ (\ref{Eq:sc_cond}) leaving the Hubbard interaction \(U\) as the only model defining parameter.

In order to solve the AIM for the GS we employ a hybrid quantum-classical variational quantum eigensolver (VQE) algorithm \cite{Peruzzo2014}. In this approach we measure the expectation value of the Hamiltonian with respect to a parameterized state \(\left|\psi(\left\{\theta_i\right\})\right>\) with the help of a quantum computer. This expectation value serves as cost function of a classical optimizer, which is used to optimize the set of parameters \(\left\{\theta_i\right\}\) until the cost function is minimal, which is then identified as the GS energy of the system according to the variational principle. The quantum state is generated by a set of unitary gates, called the ansatz, which are executed on the qubits representing the system. 

The choice of an appropriate ansatz depends on various aspects: the depth of the circuit, the number of variational parameters, symmetry properties, the amount of required computational resources, and the generalizability. A problem tailored ansatz \cite{Rungger2019, Keen2020} yields the shortest circuit with the drawback of not being generalizable. Circuits constructed by ADAPT-VQE \cite{Grimsley2019} tend to be short as well, but require a significant number of additionally quantum resources due to testing the addition of several different elements to the circuit. Hardware-efficient ansätze \cite{Kandala2017} are comparable short quantum circuits but have a large number of variational parameters and no conservation of the particle number, which is essential for the calculation of the excited states in the way we will perform them. Unitary Coupled Cluster based ansätze \cite{Anand2022} conserve the particle number and have less parameters but are significantly longer. \\
We therefore choose the particle- and \(S_z\)-number conserving ansatz presented by Gard et al.\ \cite{Gard2020}, which is a generic hardware-efficient ansatz constructed from particle-number conserving gates, as a compromise between the different aspects. 
For the GS of the two-site AIM around half-filling it is depicted in Fig.\ \ref{Fig:Quantumcircuits}. In this ansatz, 12 real parameters are required to cover the whole Hilbert space of two electrons in four spin-orbitals. The respective quantum circuit which is capable of generating any state in this 12-dimensional Hilbert space is shown in Fig.\ \ref{Fig:Quantumcircuits} (a). It makes use of a so-called A-gate, which can be decomposed into elementary single and two-qubit gates \cite{Gard2020}, as shown in Fig.\ \ref{Fig:Quantumcircuits} (d). Due to the normalization and an irrelevant global phase factor, two parameters and thus one A-gate can be removed. Further, as the Hamiltonian is symmetric under time reversal, all the \(\phi\) parameters can be set to zero. Using the \(S_z\)-symmetry of the Hamiltonian, even the \(\theta\) parameters of gates connecting both spin sectors can be fixed to zero, such that an optimization of only four parameters ($\theta_0$, $\theta_1$, $\theta_3$, and $\theta_4$ in Fig.\ \ref{Fig:Quantumcircuits} (a)) is sufficient.

For the calculation of the GF in Eq.\ \eqref{Eq:ImpLehmann} we need to compute the excited states with one electron added or removed in comparison to the GS. 
For the two-site DMFT model close to half filling the excited states are those with 1 or 3 electrons in the system, leading to four states that need to be determined in each case. As the Hamiltonian is spin-conserving and an uneven number of electrons is in these states, they can be sorted by their spin into two sets of two states. With the particle- and \(S_z\)-conserving ansatz (see Figs.\ \ref{Fig:Quantumcircuits} (b) and (c)) based on Ref.\ \cite{Gard2020} we determine the lowest energy state by VQE for \(+H\) and the highest energy state by VQE for \(-H\) in each spin subspace. So far, 8 parameters are required to cover the whole Hilbert space, which can be further reduced to 3 due to normalization, global phase and time-reversal symmetry. As one spin-sector is always completely empty (filled) no gate is required for this sector and the entanglement between both sectors. Thus, only one parameter ($\theta_0$), needs to be optimized by the VQE algorithm for these excited states. 

We quantify the quality of the results of the quantum computation by comparison with the exact results that can be obtained by an exact diagonalization (ED) method. The comparison of eigenstates is done via the fidelity defined as
\begin{align}
f = \left| \left<\Psi_{N,m}^\text{ED}|\Psi_{N,m}^\text{VQE}\right> \right|^2\label{Eq:Fidelity}
\end{align}
where we transformed the VQE state \(\left|\Psi_{N,m}^\text{VQE}\right>\) into a vector in the Fock basis and project it on the eigenvector of the Hamiltonian \(\left|\Psi_{N,m}^\text{ED}\right>\), which we obtained by ED.

With the circuits (Fig.\ \ref{Fig:Quantumcircuits}) and VQE-optimized parameters at hand, that are needed to prepare the states in Eq.\ \eqref{Eq:ImpLehmann}, we proceed to the calculation of the transfer matrix in the numerator. Exploiting the inversion of the Jordan-Wigner transformation, i.e.\ \(\prod_{i=0}^{k-1}(\sigma_i^z)\sigma_k^x = (c_k +c_k^\dagger)\), we obtain 
\begin{multline}
\left< \Psi_{N-1,m} \left| \left(\prod^{k-1}_{i=0} \sigma^{z}_i \right) \sigma^x_k \right|\Psi_{N, n}\right>\\ = \left< \Psi_{N-1,m} \left| c_k \right| \Psi_{N,n} \right>
\end{multline}
and a similar expression for the creation operator with respect to the states \(\left|\Psi_{N+1,m}\right>\), as the other term vanishes due to the fixed particle number. Thus, it is sufficient to measure only one term on the QC. That is, we combine the GS circuit, Fig.\ \ref{Fig:Quantumcircuits} (a), the circuit realization of the product of \(\sigma_i^z\) and \(\sigma_i^x\) operator and the inverse circuit of the excited state. Then, the transition probability is the probability of measuring the combined circuit in the state with all bits equal to zero.

%
%

\section{Noise models and error mitigation}\label{Sec:NoiseModels}
We have tested the implementation of our algorithm using different simulators as implemented in \texttt{QISKIT} \cite{Qiskit} as well as real quantum computing devices. We refer to the linear algebra-based
noiseless simulator as the \textit{state vector simulator}. This simulator is used to verify that the algorithm is capable to obtain the correct solution.
 
The probabilistic simulator, which takes into account the stochastic nature of the quantum mechanical measurement process is referred to as the \textit{QASM simulator} (quantum assembly language) \cite{Qiskit}. Here, the decisive quantity which determines the quality of the result is the number of shots, i.e.\ the number of repetitions of the calculation from which the probability distribution of an observable and finally its expectation value are deduced.

Furthermore, we use a noisy simulator which captures
measurement errors, single- and two-qubit
gate errors, depolarization, and thermal-relaxation errors.
We construct this \textit{fake backend simulator} by using the information
of the IBMQ Ehningen device. We also construct fake backends with rescaled error parameters.

A VQE calculation requires the choice of an optimization algorithm.
Here we use \texttt{QISKIT}'s implementation of the limited-memory Broyden-Fletcher-Goldfarb-Shannon optimization algorithm
(\texttt{L\char`_BFGS\char`_B}) for the state vector simulator and in all other cases the simultaneous perturbation stochastic approximation (\texttt{SPSA}) optimizer \cite{Kandala2017}, which is an optimizer adapted to cope with noisy data. 

When running our VQE algorithm on real hardware, we use the state-preparation-and-measurement (SPAM) error mitigation as implemented in the M3 package of \texttt{QISKIT} \cite{Nation2021}. This approach assumes a redistribution of counts only between similar bitstrings (i.e.\ differing only by one bit), such that a reduced correction matrix is sufficient, and scales at most linearly with the number of qubits.

For selected calculations we additionally used the recently published inverted-circuit zero noise extrapolation (IC-ZNE) \cite{Koenig2024}. For a ZNE the quantum circuit under consideration is run multiple times with additional randomly inserted identity operations. These identity operations are constructed by an appropriate repetition of the most noisy gates, in our case the CNOT-gates, which increases the noise level in the circuit. Measuring the expectation values corresponding to these extended circuits as funtion of the added noise allows to extrapolate to zero noise. While in the standard ZNE approach the number of gate repetitions is used to quantify the amount of noise, in IC-ZNE it is the actual error strength of the circuit. This error strength is defined by the probability to measure the all-zero bit-string when the noise amplified circuit is combined with its inverse. Further, randomized compiling \cite{Wallman2016} was used in order to un-bias the noise in the noise-enhanced circuits.
For our IC-ZNE calculations we used 4 amplification rates with 10 randomized compilations at 1000 shots each and a linear extrapolation.

%
%

\section{Results}\label{Sec:Results}
In this section we proceed along the sequence of different steps in the DMFT self-consistency loop and discuss the results obtained using different levels of noise as outlined above.
This, in detail, comprises the determination of the different states and their energies by VQE, the calculation of the GF in the Lehmann representation and thus the transition probabilities, the determination of the quasi-particle weight and finally the full self-consistent solution.


\subsection{Energy Spectrum}\label{Sec:Energies}

\begin{figure}
\setlength{\unitlength}{1mm}
\begin{center}
\begin{picture}(85,93)(0,0)
\put(0,0) {\includegraphics[width=\columnwidth]{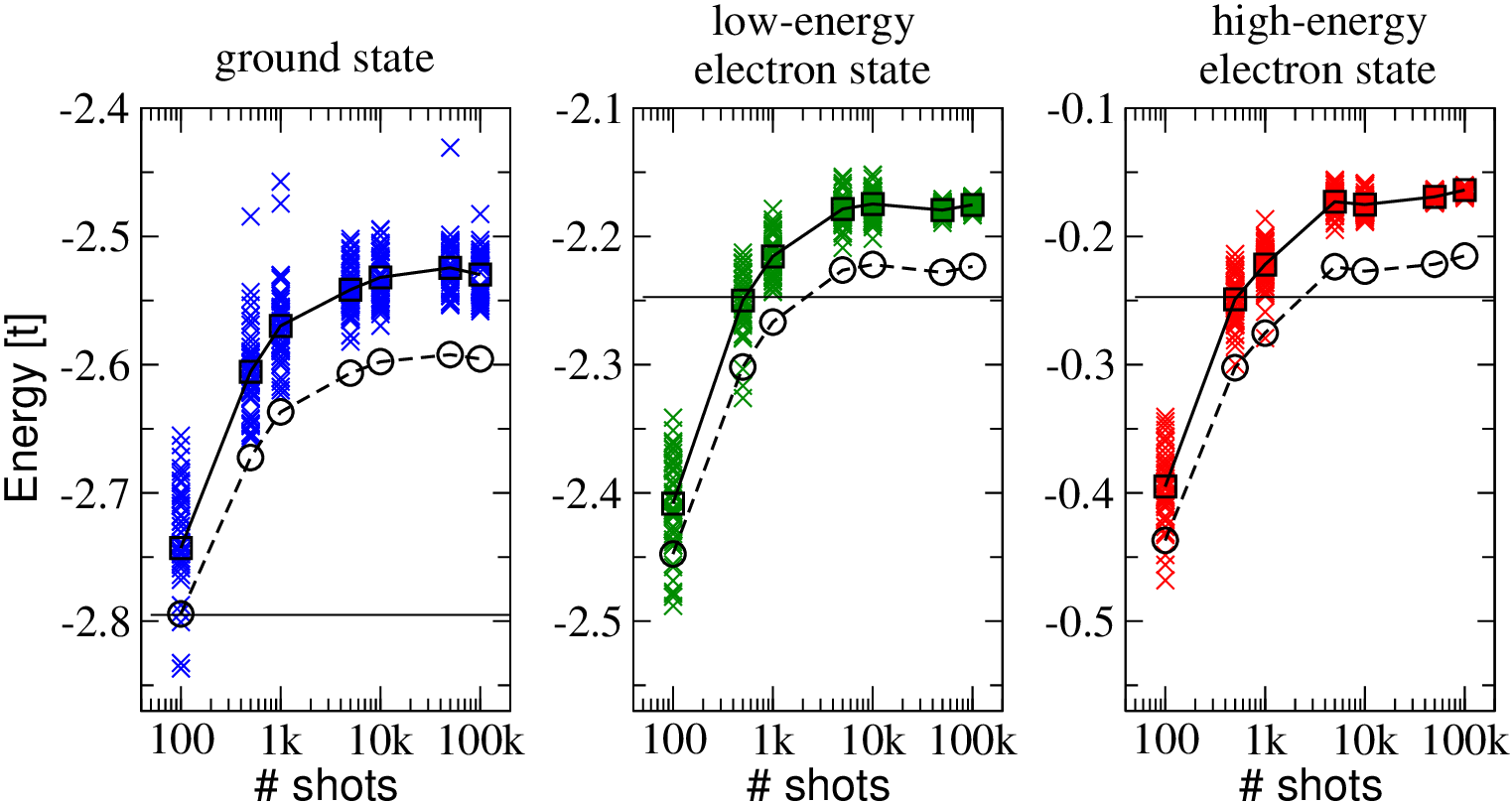}}
\put(0,48){\includegraphics[width=\columnwidth]{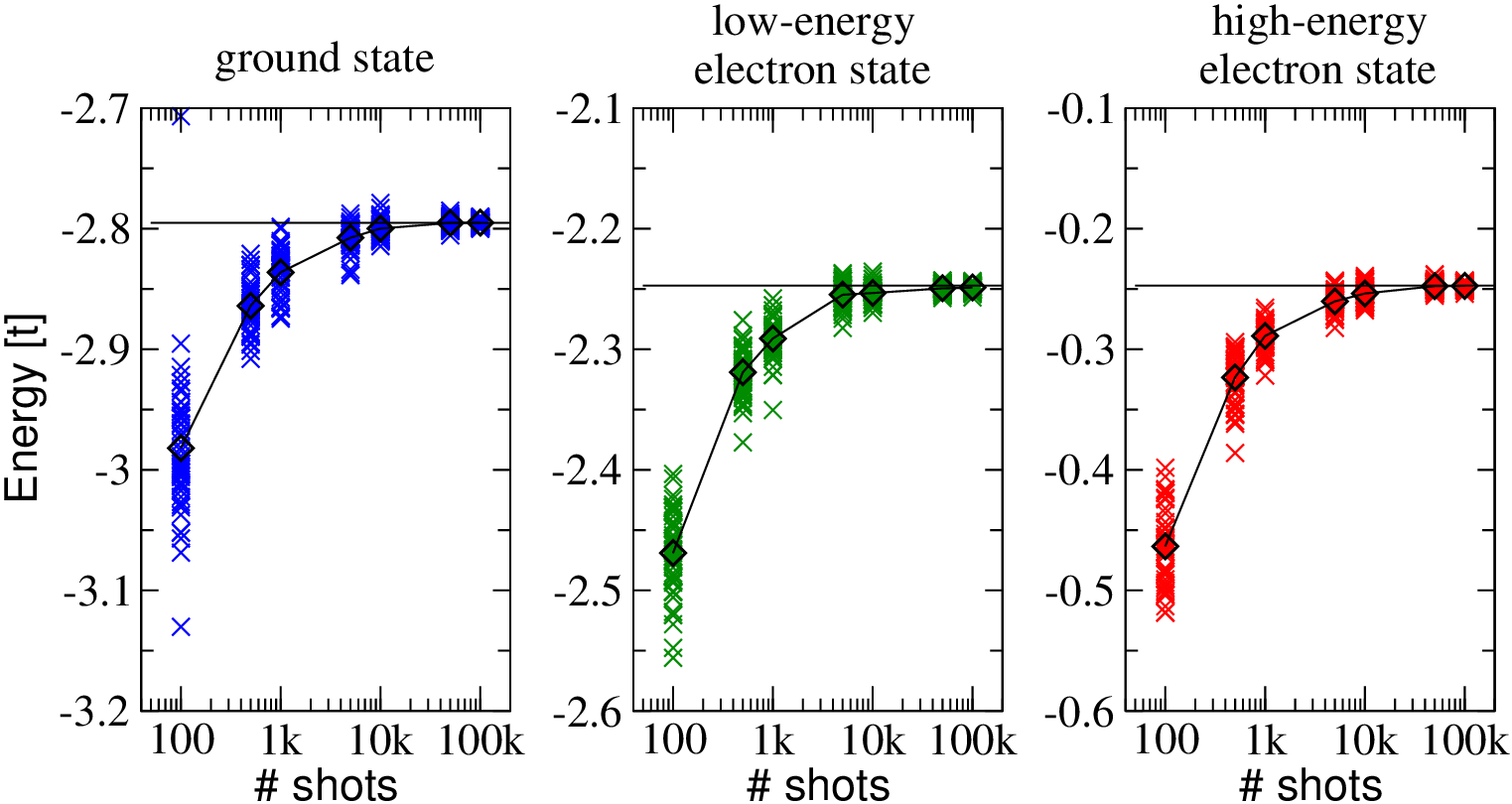}}
\put(0,43){(b)}
\put(0,90){(a)}
\end{picture}
\end{center}
\caption{Convergence of energies based on VQE as a function of the number of shots, using (a) the QASM simulator and (b) a fake backend. Shown are results for the ground state with \(N=2\), and the low- and high-energy electron  spin-down states $N=3$, $S_z=-1$). The diamond symbols connected by lines represent the mean calculated from 50 runs, while the cross symbols refer to the result of individual runs. Horizontal solid black lines indicate the exact eigenenergies. 
Post-processing of the fake backend results by M3 error mitigation yields improved mean energy values shown by circle symbols connected by dashed lines.
\label{Fig:shot_conv}}
\end{figure}

We consider the two-site DMFT model at half-filling with \(\varepsilon_d=0, \varepsilon_c=\nicefrac{U}{2}=\mu\) and \(V\) according to Eq.\ \eqref{Eq:V_halffilled}. In the first step, we determine the eigenstates (eigenenergies and eigenvectors) required for the Lehmann representation of the GF by VQE optimization using ansatz circuits with \(N=2\) and \(S_z=0\) for the global GS and for all the states in the \(S_z=\pm 1\) subspaces for \(N=1\) and \(N=3\) as described in Sec.\ \ref{Sec:QC}. 

The eigenenergies obtained by using our QC approach with the state vector simulator and the \texttt{L\char`_BFGS\char`_B} optimizer agree with those from ED of the Hamiltonian up to \(10^{-8}\ t\). In addition, the states obtained by VQE have a fidelity close to 1 with respect to the exact eigenstates. Thus, the chosen ansatz states with the parameters obtained from VQE are well suited to reliably obtain the states of the system required for the GF.

\subsubsection{QASM Simulations}
Next, we perform the VQE optimization employing the probabilistic QASM simulator with a finite number of shots. Here we use the \texttt{SPSA} optimizer. By increasing the number of shots, the measured energies are expected to become more and more precise, as the stochastic accuracy grows with the number of shots \(N\) by \(\propto \sqrt{N}\), so that the optimizer should improve in finding the correct states. In the limit of an infinite number of shots we should, thus, recover the exact results as obtained by ED or with the state vector simulator. 
We performed the VQE optimization 50 times for each chosen number of shots for all the states. This allows to average out the effect of the stochastic noise and the stochastic nature of the \texttt{SPSA} optimizer.

Figure \ref{Fig:shot_conv} (a) shows the evolution of the minimal measured eigenvalues in the optimization with the shot number for the GS, one low-energy electron, and one high-energy electron state. When the number of shots is increased, the accuracy of the results is improved and the spread of the results is significantly reduced. The surprisingly small values for a small number of shots, which are even below the exact GS of the system, result from the statistical fluctuations and our choice of criteria to determine the best parameters. The SPSA optimizer as implemented in \texttt{QISKIT} returns the last parameters accepted, even if at some point in the optimization history smaller cost values were obtained. Therefore, we tracked all intermediate results and then took the overall lowest cost values with their corresponding parameters. For a small number of shots, strong statistical fluctuations lead to eigenvalues, which might be even below that of the GS. As we collected about 200 values for the cost function, we have a large chance to choose such a statistical outlyer as our optimal value, so that we sample the lowest branch of the curve which is proportional to \(\sqrt{N}\) for the expectation values. When the number of shots is increased, the statistical fluctuation is reduced, and thus we obtain expectation values matching the GS. From this analysis we conclude that at least \(10\, k\) shots are required to obtain reasonable and reliable results. 

\subsubsection{Fake Backend Simulations}
In order to investigate how much the errors of a real backend influence the results of successful QC calculations, we performed the same VQE simulations (SPSA optimizer, 200 iterations) with a fake backend based on the real error-rate data available for the IBMQ Ehningen \cite{ErrorValues}. In Fig.\  \ref{Fig:shot_conv} (b) we show the resulting energies for 50 VQE runs per shot number. Their mean values (diamond symbols) show the same trend as those of the \texttt{QASM} result, however, they converge to values deviating from the exact ones by about \(10\,\%\) and \(5\,\%\) for the GS and the electron state, respectively, due to the simulated device errors. In order to differentiate the influence of SPAM errors from gate and decoherence errors, we have applied M3 error mitigation \cite{Nation2021} to these fake backend results. The average values of the mitigated data are shown as circle symbols in Fig.\  \ref{Fig:shot_conv} (b). They are improved by about $0.7t$ and $0.5t$ compared to the unmitigated raw data, for the GS and the excited states, respectively.

The electron state circuits are short and contain only three  CNOT gates. Thus, the decoherence and the gate errors have a weak influence on the results in comparison to SPAM errors. The length of the GS circuit is still relatively short, only three times longer than the minimal length of the excited state circuits, such that decoherence errors are negligible. However, the GS circuit contains 15 CNOT gates and thus 
it is significantly more affected by the two-qubit gate errors, leading to the larger error in the resulting expectation value. These dependencies are supported by our calculations: the addition of M3 error mitigation to the fake-backend simulations corrects the expectation values of the excited states almost perfectly, while it hardly improves the GS results. Therefore, we conclude that the SPAM errors, which are dominant in short circuits, are well reduced by the M3 error mitigation, and the main limitation for the real-backend implementation of our calculations is the magnitude of the CNOT errors.

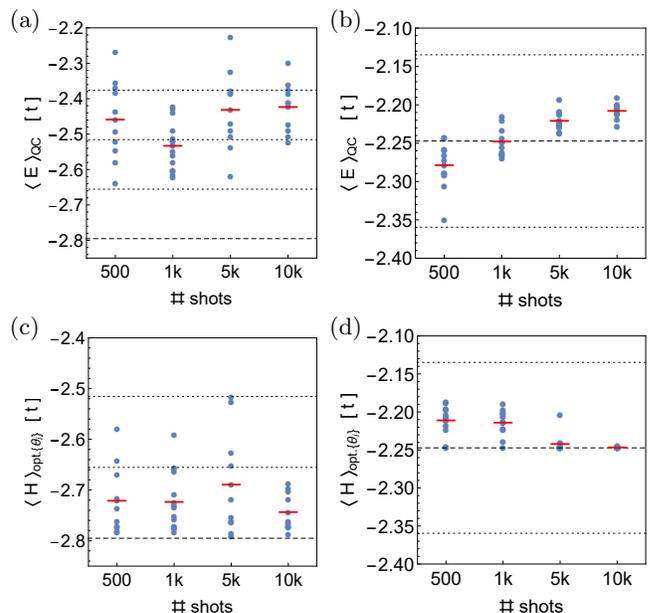
\begin{figure}
\begin{center}
\setlength{\unitlength}{1mm}
\begin{picture}(85,81)(0,0)
	\put(2,41){\includegraphics[width=0.45\columnwidth,draft=false]{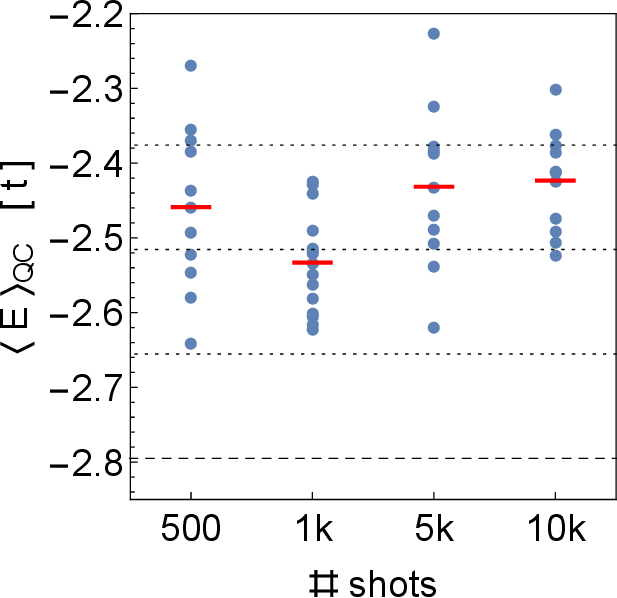}}
	\put(2,0){\includegraphics[width=0.45\columnwidth,draft=false]{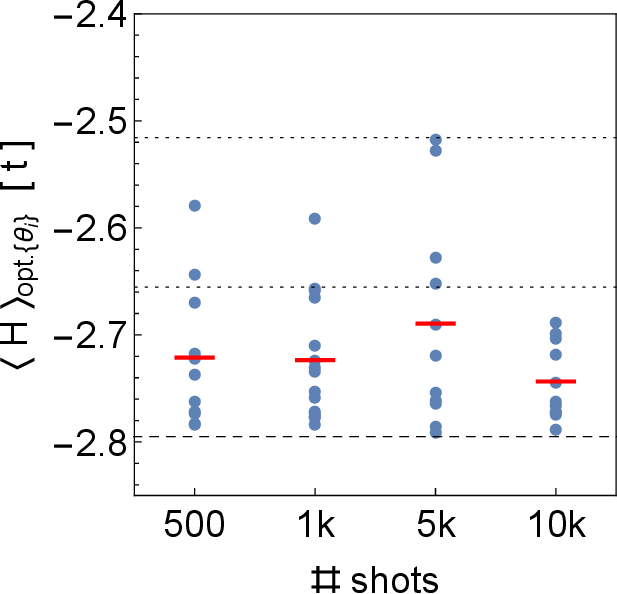}}
	\put(44,40.7){\includegraphics[width=0.47\columnwidth,draft=false]{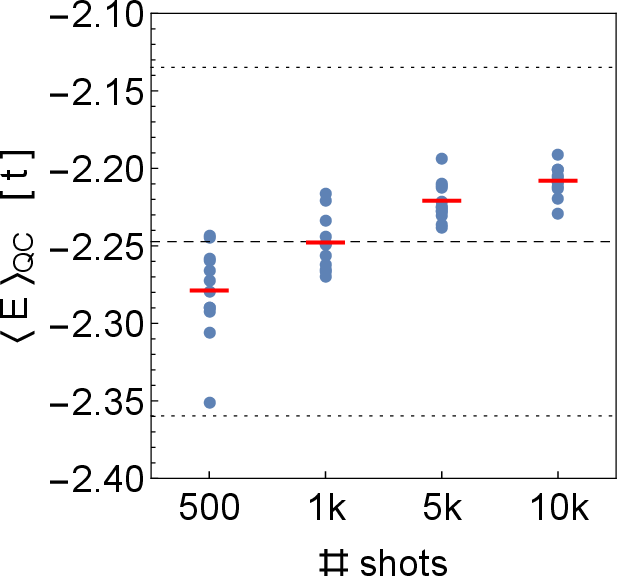}}
	\put(44,0){\includegraphics[width=0.47\columnwidth,draft=false]{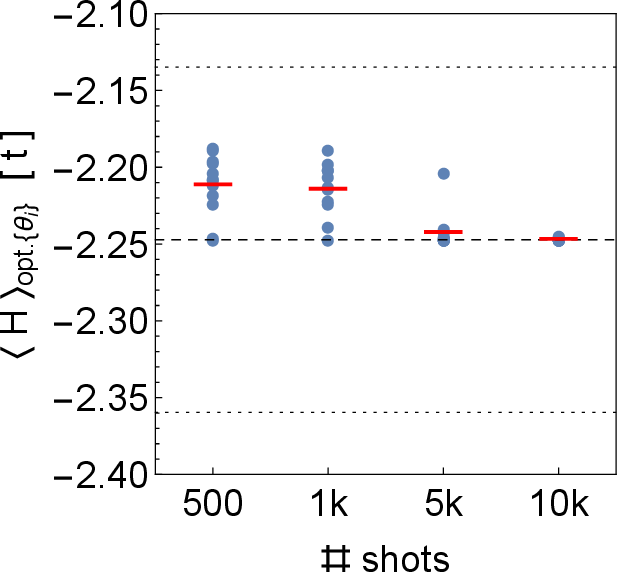}}
	\put(0,78){(a)}
	\put(42.5,78){(b)}
	\put(0,37){(c)}
	\put(42.5,37){(d)}
\end{picture}
\caption{Expectation values for the energy of (a) the ground state and (b) the low-energy electron spin-down state obtained by independent VQE calculations on IBMQ Ehningen for different numbers of shots. The blue points correspond to ten individual runs for each shot number and the red lines to their mean value. In (c) and (d) we show the exact energy expectation values corresponding to the optimal parameters found in the VQE runs of (a) and (b), respectively. The dashed black horizontal lines indicate the exact value and the dotted lines deviations in steps of \(5\,\%\). }\label{Fig:shot_ehningen}
\end{center}
\end{figure}

\subsubsection{Real Backend Calculations}
We compare the results of a simulated quantum backend with results of VQE calculations on the real backend IBMQ Ehningen. On the real backend we use the same setting (SPSA optimizer, 200 iterations, M3 error mitigation in each VQE step) but vary the number of shots only up to \(10 k\) as this appeared sufficient by the previous simulations. In Figs.\ \ref{Fig:shot_ehningen} (a) and (b) we show the results of 10 individual VQE calculations for each number of shots (blue dots) and their mean (red bar) for the GS and the low-energy electron spin-down state, respectively. Here, similar to the fake backend results, we observe an overall increase in the expectation value with the number of shots. This is due to the selection of the overall lowest-energy state encountered during the SPSA optimization, leading to a preference of the lowest outliers, as discussed for the QASM case above. However, the spread of expectation values is not reduced between \(1 k\) and \(10 k\) shots in both cases.
In addition we note, that the absolute values obtained from the real backend IBMQ Ehningen and those from the fake backend agree sufficiently well, regarding the natural statistical fluctuations within \(0.1\,t\) for both states. This indicates that the fake backend mimics the behavior of the real backend reasonably well
and that our conclusions on the relevance of the different sources of errors from the previous section hold.
However, the calculations on the real backend on average yield errors that are increased by a few percent compared to the fake backend simulations. This reflects that some aspects of the real hardware are not accounted for in the simulated device. This is particularly true for the so-called cross-talk errors,\cite{Ketterer2023} which refer to the undesirable influence that a gate operation on a given qubit (or pair of qubits) has on its physically neighboring qubits.  

To analyze the accuracy of the state obtained by VQE on a real backend, we calculate the exact expectation values corresponding to the optimal parameters determined in the respective VQE runs. The resulting expectation values, shown in Figs.\ \ref{Fig:shot_ehningen} (c) and (d), are significantly improved compared to the evaluation on the quantum hardware in both cases. For the electron state, we observe a very good convergence to the exact value with \(10\,k\) shots. This indicates that the presence of noise is leading to a systematic offset in the one-dimensional energy landscape during the VQE optimization. For the GS the exact expectation values corresponding to the obtained optimal parameters have, in most cases, an error of less than \(5\, \%\). Thus, in the presence of CNOT errors and statistical noise, although the optimizer is not able to find the global minimum in the 4-parametric energy landscape, it leads at least to a state close to it. Its quality (i.e. the quality of the optimized variational parameters) is actually much better, than we would expect based on the expectation value obtained directly from the device that yields errors of more than \(12\, \%\) on average.  

Based on the results above, we have to perform our calculations with at least \(10\,k\) shots in order to reduce shot noise to a tolerable level. While M3 mitigates SPAM errors very well, the CNOT errors are a significant threat to obtain accurate results on the real hardware. We will further investigate their effect in Sec.\ \ref{Sec:Mott_MIT}. However, in order to go through the steps of our DMFT algorithm, we first discuss the computation of the GF with respect to stochastic errors in the following section.


\subsection{Green's function}\label{Sec:GF}

\begin{figure}
\begin{center}
\includegraphics[width=\columnwidth]{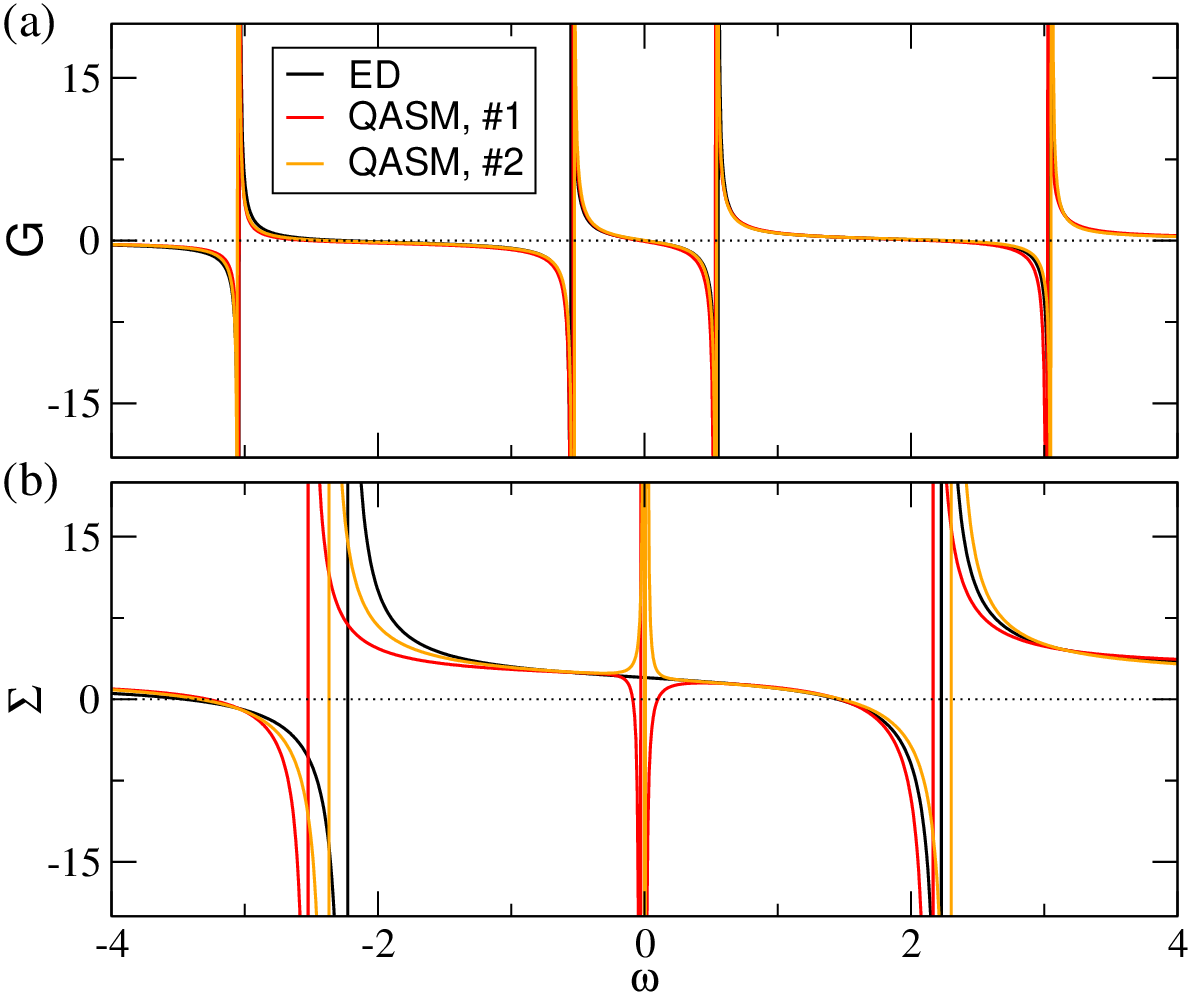}
\caption{Impurity GF (a) and self-energy (b) for the half-filled two-site DMFT model with \(U=4\,t\). The exact diagonalization (ED) and state vector VQE result are in perfect agreement and we only show the ED result. The two runs using the QASM simulator with \(10\,k\) shots for the VQE optimisation and transition rates illustrate the difficulties arising from inaccurate QC results.}\label{Fig:GF_Selfene}
\end{center}
\end{figure}

According to Eq.\ \eqref{Eq:ImpLehmann}, the GF of the two-site DMFT model has a four-peak structure where the peaks at negative frequencies originate from hole excitations and those at positive frequencies from electron excitations. Their positions are determined by the corresponding energies relative to the GS energy and their shapes by the transition rates. Due to the electron-hole symmetry of our model at half-filling, the GF is symmetric with respect to \(\omega=0\). As the inverse GF and its free counterpart appear in the Dyson equation \eqref{Eq:Dyson}, the self-energy has poles at energies where only one of the two GFs is zero as the divergences cancel otherwise. 
While the full GF has three zeros, the free GF has only one. As this coincides with the zero transition of the full GF at zero-frequency, the remaining two zeros yield two singularities in the self-energy.

When energies obtained by VQE calculations with the state vector simulator are used to construct the impurity GF and the self-energy for \(U=4\,t\), a perfect agreement is achieved with the exact quantities obtained from ED (cf.\ Fig.\ \ref{Fig:GF_Selfene}). This could be expected by the high precision in energies and the high fidelities, compared to the exact results, as discussed in the previous section. Thus, the high fidelity leads to transition rates which are also in very good agreement with the exact values. To obtain the derivative of the self-energy at \(\omega=0\) independent of small numerical fluctuations we use the first coefficient of a linear fit to its curve in the proximity to this frequency. Thereby we obtain the quasi-particle weight by Eq.\ \eqref{Eq:QP_weight} as \(z=0.5552\) which is very close to the exact value of \(z_\text{ex}=0.5556\).

\subsubsection{QASM Simulations}
When we use the results from stochastic QASM simulations, the constructed GFs differ only slightly for at least \(10\,k\) shots while there are larger differences observable in the self-energy (see Fig.\ \ref{Fig:GF_Selfene}). This results from the imprecision in the VQE calculation of the eigenstates. While deviations in the eigenenergies shift the poles of the GF, the eigenvectors influence the transition rates, which are relevant for the curvature of the GF. Furthermore, the errors on the eigenstates lift the electron-hole symmetry. In total, the zeroes in the GF are shifted, resulting in corresponding shifts of the peaks in the self-energy. Most crucially,
the coincidence with the zeroes of the free GF is lost. This results in a two-peak structure close to \(\omega = 0\) which prevents us from taking the derivative of the self-energy (cf.\ Eq.\ \eqref{Eq:QP_weight}) at this point. In addition, the shift of the other peaks has an influence on the curvature at that point, too. Figure\ \ref{Fig:GF_Selfene} exemplarily shows this for two runs with \(10\,k\) shots. 
We observed that even \(100\,k\) shots are not sufficient to get rid of this artificial two-peak structure at \(\omega=0\).

A possible approach to overcome these unphysical quasi-particle signatures follows a regularization procedure as first suggested in Ref.\ \cite{Rungger2019}. Here, the transition rates and energies are corrected such that the zeroes and the derivative of the impurity GF matches those of the free GF at the bath energies \(\varepsilon_c\). However, we do not follow this route here, as for some test cases the correction terms were significantly larger than the measured values and sometimes lead to unphysical results with negative transition rates.

\subsubsection{The tan-fit-Approach}
We propose to take a more pragmatic route and fit a function of the form \(a\cdot \tan(\omega) + b\cdot \omega + c\) to the self-energy in the region between the two physical peaks. This function was chosen as the Taylor expansion around \(\omega=0\) of the analytic self-energy consists only of odd-integer exponents similar to the expansion of the tangent. We refer to it in the following as the \(\tan\)-fit approach.
Due to the origin of the artificial peaks, we first search the zero of the free GF closest to \(\omega = 0\), denoted by \(\omega_0^{0}\)  (in the half-filled case \(\omega_0^0 = 0\)), and the zero of the full GF closest to it (\(\omega_0\)). The first zero of the full GF on the negative and positive frequency axis which is not \(\omega_0\), denoted by \(\omega_{-}\) and \(\omega_{+}\), define the position of the physical peaks. We take \([\omega_-, \omega_+]\) as our fitting interval and exclude the interval \([\omega_0^{0}, \omega_0]\). The quasi-particle weights can now be obtained from the derivative of the fitting function at \(\omega=0\). We test this approach for the exact self-energy and obtain the quasi-particle weight \(0.542\) that is reasonably close to the exact value of \(z_\text{ex} = 0.555\). 

\begin{figure}
\includegraphics[width=\columnwidth,draft=false]{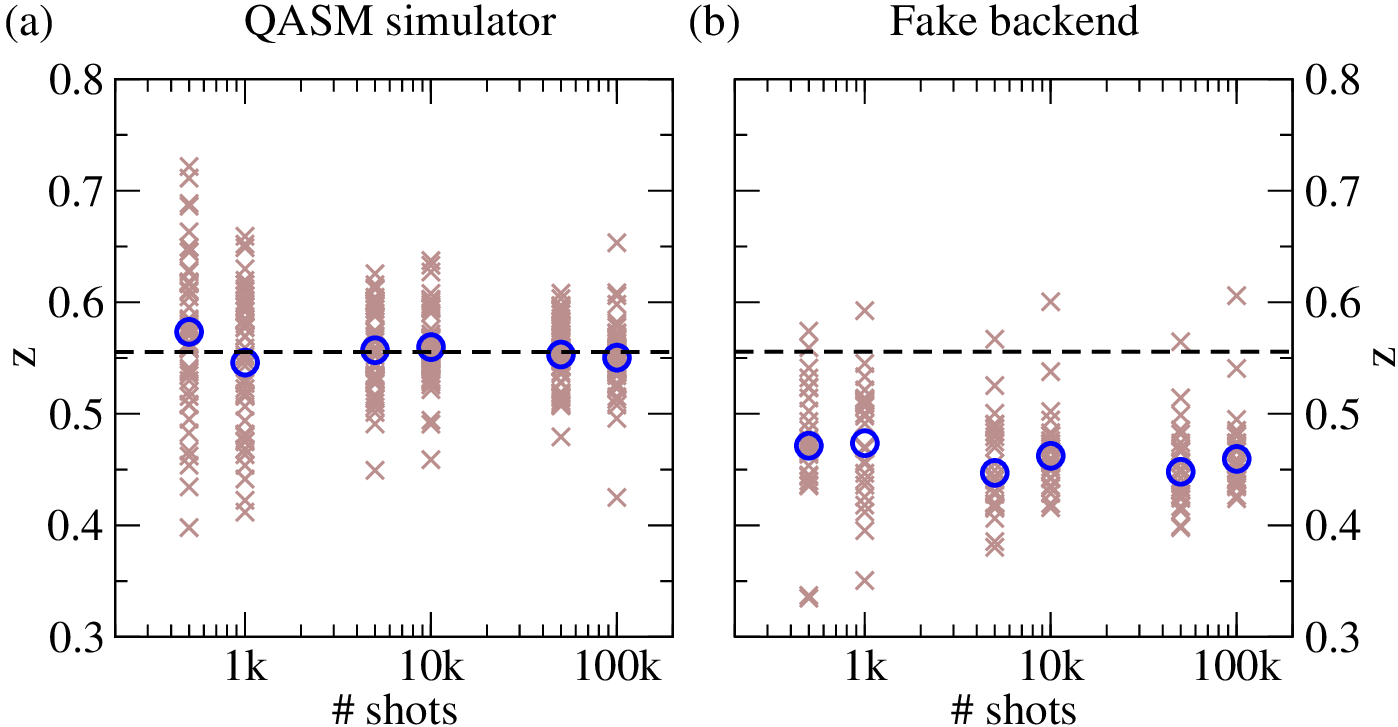}
\caption{The quasi-particle weight \(z\) obtained via the \(\tan\)-fit procedure applied to the self-energies as function of the number of shots. The self-energy is calculated for the two-site DMFT system at half-filling and \(U=4\,t\) with the analytic exact \(V\) with (a) probabilistic QASM simulations and (b) a fake backend. Due to the stochastic nature, we performed 50 runs (gray crosses) and show their arithmetic mean (blue circles). The exact value for this parameter setting is marked by the dashed black lines.}
\label{Fig:zvsshot}
\end{figure}

In Fig.\ \ref{Fig:zvsshot} (a) we display the quasi-particle weights obtained by this \(\tan\)-fit procedure applied to the self-energies obtained with a finite number of shots. The determined quasi-particle weights are spread around the exact values. When the number of shots is increased, the spread is reduced, which agrees well with the more accurate results in this case described above. However, even for \(100\,k\) shots the results of individual runs differ from the exact value by up to \(10\%\) due to the stochastic shot noise.

\subsubsection{Fake Backend Simulations}
When we use a fake backend with a noise model based on data taken from the real system IBMQ Ehningen \cite{ErrorValues}, we obtain the quasi-particle weights displayed in Fig.\ \ref{Fig:zvsshot} (b) from the \(\tan\)-fit approach. While the spread of the individual results reduces with a larger number of shots, the quasi-particle weight is always about \(15\%\) smaller than the exact one. Tracing back this systematic deviation, we reconsider the construction of the GF. 
As the eigenvalues of the GS are overestimated by about \(0.27t\) and those of the excited states by approximately \(\approx 0.08t\), the frequencies determining the peaks of the GF are underestimated. This creates a tendency of the zero-crossings to be located closer to each other compared to the exact analytic calculation. Therefore, the peaks in the self-energy are located too close to each other and the \(\tan\)-fit procedure results in an overestimation of the slope and thus an underestimation of the quasi-particle weight.

\begin{figure}[]
\setlength{\unitlength}{1mm}
\begin{center}
\begin{picture}(85,200)(0,0)
\put(5,100) {\includegraphics[width=0.88\columnwidth,draft=false]{Fig7a_QC_GF_new.epsi}}
\put(5,0){\includegraphics[width=0.88\columnwidth,draft=false]{Fig7b_QC_Sigma_new.epsi}}
\put(0,97){(b)}
\put(0,197){(a)}
\end{picture}
\caption{(a) Impurity GF and (b) self-energy at half-filling at the self-consistent condition with \(U=4\,t\) obtained from five experiments on IBMQ Ehningen compared to the exact result (ED). The GF is obtained from VQE calculations for the GS and excited states employing the \texttt{SPSA} optimizer, \(10\,k\) shots and M3 readout error mitigation. The GF and self-energy based on the bare expectation values (red lines) is significantly improved by an IC-ZNE mitigation for the ground state (blue lines).}\label{Fig:QC_results}
\end{center}
\end{figure}

\subsubsection{Real Backend Calculations}

We performed the full set of VQE calculations and transition-rate measurements on the state-of-the-art QC IBMQ Ehningen 5 times with \(10\,k\) shots. The resulting GFs and self-energies are displayed in Fig.\ \ref{Fig:QC_results} (solid red lines). The GFs in plot (a) show the shift of the peaks towards \(\omega=0\) due to the imprecision in the energies obtained by VQE and, thus, of the zero-transition. The latter is also affected by the transition rates, whose measurement results in  errors of up to \(25\,\%\) with respect to an analytic evaluation. Figure \ref{Fig:QC_results} (b) shows the effect of these inaccuracies on the self-energy. The single-peak or double-peak structures close to \(\omega=0\) result from the missed cancellation of zero-transitions in the free and impurity GF, as discussed above for the QASM simulations. 
The peaks of the self-energy on the other hand are shifted away from their original position. As the \(\tan\)-fit performs a fit between these two peaks, their position has an influence on the slope at \(\omega=0\). Performing this fit, we obtain the values listed in Tab.\ \ref{Tab:QC_z}, where we see a surprisingly good agreement with the exact value when the shift is only small as in the apparently good QC-run \# 2. However, the obtained value significantly deviates for the particularly bad run QC \# 3. The results of the real backend qualitatively are in line with those obtained from the fake backend presented before, cf.\ Fig.\ \ref{Fig:zvsshot} (b).

As the main error affects the expectation value of the GS, we performed a post-processing IC-ZNE for those states as obtained by VQE. By replacing the directly measured expectation value with the extrapolated one, the GF and self-energies displayed by blue lines in Fig.\ \ref{Fig:QC_results} were obtained. It appears, that the GF can be significantly improved, especially \mbox{QC \# 1} and \mbox{QC \# 2} match the exact one very well. This also transfers to a good agreement for the self-energy. For the particularly bad run \mbox{QC \# 3} only a marginal improvement by IC-ZNE is achieved. Nevertheless, for all runs our \(\tan\)-fit approach leads to a significantly better agreement for the \(z\)-values when using the functions based on the IC-ZNE ground-state energy (cf.\ Tab.\ \ref{Tab:QC_z}).

\begin{table}
\begin{tabular}{l|ccccc|c}
run \#         &   1   &   2   &   3   &   4   &   5   & average \\ \hline
\(z\)          & 0.507 & 0.528 & 0.387 & 0.460 & 0.484 & 0.473 \\ \hline
\(z\) (IC-ZNE) & 0.566 & 0.625 & 0.431 & 0.572 & 0.545 & 0.548
\end{tabular}
\caption{Quasiparticle weight \(z\) obtained by the \(\tan\)-fit approach applied to the self energies obtained from 5 independent calculations on the IBMQ Ehningen. The first row corresponds to the plain data (c.f. red lines in Fig. 7) while for the second row a post-processing IC-ZNE was carried out for the ground state (c.f. blue lines in Fig. 7) 
}\label{Tab:QC_z}
\end{table}


\subsection{Resources}\label{Sec:Scaling}
Based on the previously reported calculations we estimate the computational costs for a complete self-consistent calculation of the GF on a QC using the Lehmann representation. The Hamiltonian of the model consists of three qubit-wise non-commuting sets of Pauli strings. Thus, for one expectation value we need to run the quantum circuits with measurements in each corresponding basis (cf.\ Fig.\ \ref{Fig:Quantumcircuits}). Therefore the number of circuits to be performed is three times the number of shots. 

We estimate the respective run time of one circuit by using approximate execution times from current IBMQ devices. Two-qubit gates require about \(400\, ns\) and the readout and measurement takes \(700\, ns\) while single qubit gates are sufficiently fast or even virtual such that we do not take them into consideration. Assuming parallel execution of CNOT gates and readout, the execution of the GS quantum circuit consisting of 9 layers of CNOT gates requires about \(4.3\, \mu s\).  As currently the state preparation is achieved by thermalization of the system, between two circuit executions a waiting time is required, which is of the same order as the system's \(T_1\) time. As this is about \(150\, \mu s\) for the IBM systems, this is the dominating time scale for the calculations we consider in the following. Therefore, one expectation value with \(10\, k\) shots requires about \(4.5\,s\) instead of \(0.13\,s\) for the plain circuits. In practice, however, we observed average run times of about \(20\, s\) using the IBM runtime sampler, indicating an overhead that we did not address in our considerations.

The hybrid algorithm presented in Sec.\ \ref{Sec:DMFT} requires one VQE calculation for the GS and eight VQE calculations for the excited states in the case of the two-site DMFT model in each step of the self-consistency loop, if no symmetries are exploited or present. For the VQE runs with stochastic noise and the SPSA optimizer (cf. Sec.\ \ref{Sec:Energies}) we observed a convergence after 170 to 300 iterations for the GS and after about 100 iterations for the excited states. However, on a real device our convergence criterion did not lead to an end of the optimization process, such that 250 cost function evaluations were performed in total. The measurement of these 250 eigenvalues thus requires about 90 minutes on the real device, while the plain circuit execution requires only about half a minute (32.5 s).

The self-consistent DMFT loop for the optimization of the hybridization requires in the best case, using the state vector simulator and BFGS optimizer, 10 to 30 iterations, depending on the interaction strength, until self-consistency is reached up to an accuracy of \(5\cdot 10^{-3}t\). Thus, a total of 90 to 270 VQE optimizations are required, which in the best case would lead to a computation time of \(135\, h\) on a real device, while the pure circuit execution time (without state preparation) is less than an hour (49 min). As we do not have access to sufficient compute time on quantum computers, we investigate our self-consistent hybrid algorithm on simulated or fake backends (cf.\ Sec.\ \ref{Sec:Self_cons}) in the following.


\subsection{Self-Consistency Loop}\label{Sec:Self_cons}

\begin{figure}
\includegraphics[width=\columnwidth]{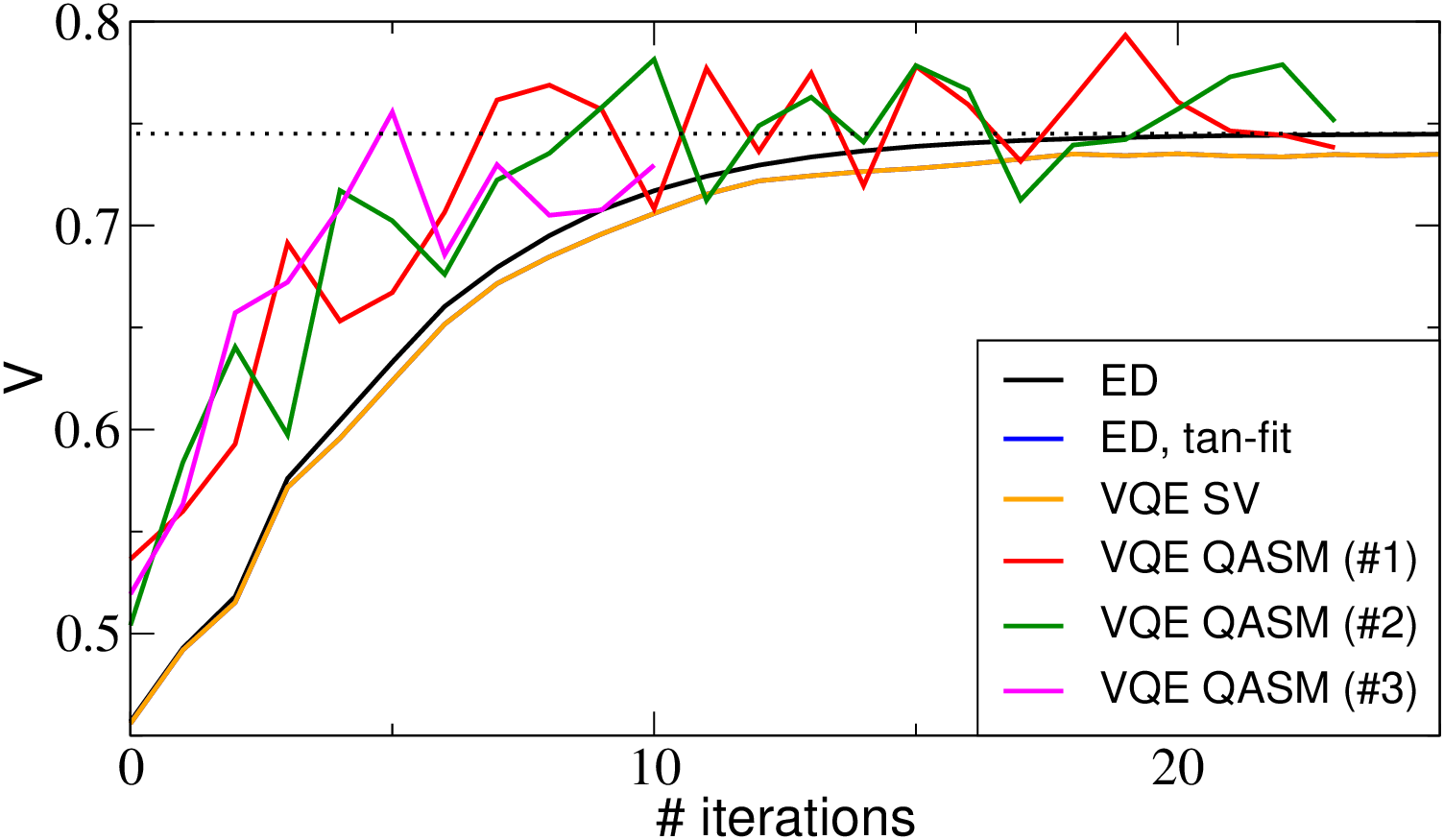}
\caption{Convergence of the hybridization \(V\) in the DMFT self-consistency loop at half filling, \(U=4t\) and initial value \(V_\text{init}=0.4\, t\) using different variants to solve the AIM. While for the curve "ED" the numerical derivative at \(\omega=0\) was taken, the \(\tan\)-fit approach was used in all other cases. The VQE results were obtained with either the state vector simulator and \texttt{L\char`_BFGS\char`_B} optimizer (SV)  or with the QASM simulator and SPSA optimizer using \(10\, k\) shots. } \label{Fig:convergence}
\end{figure}

We use our VQE based approach to obtain a self-consistent solution of the two-site DMFT model at half filling, i.e.\ we perform the inner optimization loop as displayed in Fig.\ \ref{Fig:DMFT_loop}. Figure \ref{Fig:convergence} demonstrates the convergence of the hybridization strength at half filling and \(U=4\,t\) for different solvers starting from an initial value \(V_\text{init} = 0.4\, t\) towards the exact value of \(V_\text{exact} = 0.745\, t\). For ED, the positions of zeroes of both GFs perfectly agree, such that the respective divergences in the self energy cancel and the slope of $\Sigma$ at \(\omega = 0\) can be numerically determined accurately. Thus, \(V\) smoothly converges to the exact value with a threshold of \(10^{-3}\,t\). When we use this solver but apply our \(\tan\)-fit approach to obtain the quasi-particle weight, the hybridization converges in the same way to a value of \(0.734\,t\) which matches the exact value within \(1.5\%\). The convergence is the same in the case of our quantum circuit based approach with the state vector simulator and the \texttt{L\char`_BFGS\char`_B} optimizer, such that both graphs overlap in Fig.\ \ref{Fig:convergence}. 

The convergence for three different runs with the QASM simulator using \(10\,k\) shots is also shown in Fig.\ \ref{Fig:convergence}. Although the values fluctuate strongly, the curve converges towards the correct values for all runs. This behavior is achieved by our mixing scheme based on weighted means, as described after the iteration loop in section \ref{Sec:DMFT}. 


\subsection{Mott metal-insulator transition}\label{Sec:Mott_MIT}
\begin{figure}
\center
\includegraphics[width=\columnwidth]{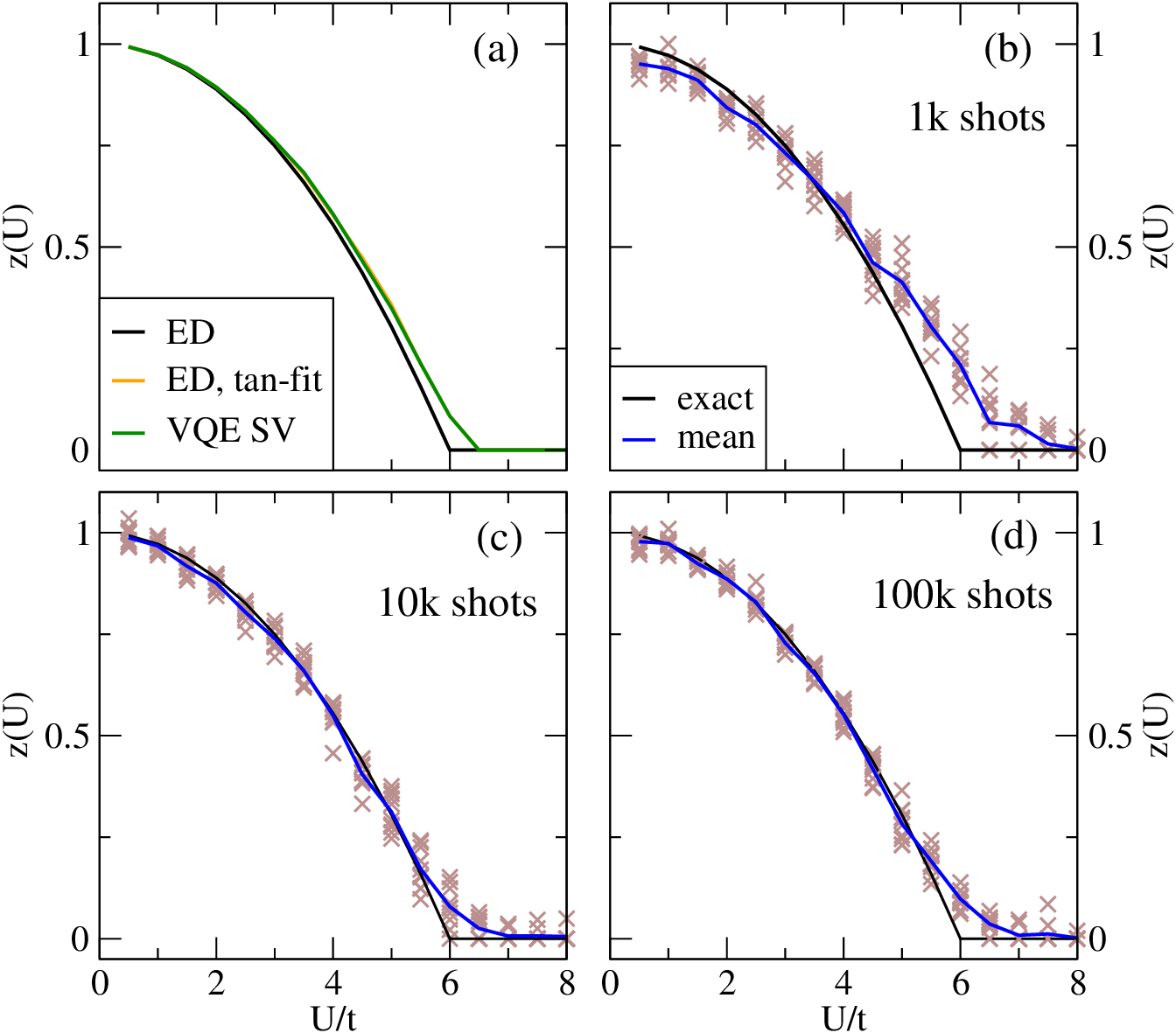}
\caption{(a): Dependence of the quasi-particle weight \(z\) on the interaction strength $U$ obtained from a self-consistent DMFT cycle when using exact diagonalization without (ED) and with the \(\tan\)-fit (ED + tan-fit) and the VQE solver with the state vector simulator (VQE SV). (b)-(d): Same as (a) but with the QASM simulator and \(1\,k\), \(10\,k\) and \(100\,k\) shots, respectively. The crosses mark the final values of 10 individual runs, while the blue lines mark their arithmetic mean values in comparison to the analytic result (black line). 
} \label{Fig:MIT_half_filling}
\end{figure}

Motivated by these results we investigate the Mott metal-insulator transition of the two-site DMFT model at half filling. Figure \ref{Fig:MIT_half_filling} shows how the quasi-particle weight \(z\) evolves with the interaction strength \(U\) for the different approaches investigated from finite values in the metallic phase for \(U<U_c=6\,t\) to the Mott-insulating phase for \(U>U_c\). 

Overall, ED in combination with the \(\tan\)-fit approach reproduces the quasi-particle weight over the whole \(U\)-range very well in comparison to the analytic result \cite{Potthoff2001}. Although it tends to overestimate \(z\) in the metallic region close to the Mott transition, it still reproduces a sharp transition, however at a slightly larger \(U\) value. Simulations using the state vector simulator exactly reproduce the ED results. 

With the QASM simulator we performed 10 independent self-consistent DMFT simulations for each \(U\)-value  using either \(1\,k\), \(10\,k\), or \(100\,k\) shots. In Fig.\ \ref{Fig:MIT_half_filling} (b)-(d), we show both, the individual results (crosses) as well as the mean result of all these runs (blue line) together with the analytic result (black line). With \(1\,k\) shots we were able to reproduce the curve only for \(U<4\,t\) while we failed to obtain the Mott transition. Calculations with \(10\, k\) or \(100\,k\) shots resulted in a much better agreement over the whole \(U\)-range and (for most runs) a vanishing \(z\) for the insulating phase. However, here we get a smooth transition instead of a sharp edge at \(U=6\,t\). Despite this most challenging feature our pragmatic approach can overcome the limitations of noisy data obtained on a NISQ computer.

\begin{figure}
\includegraphics[width = \columnwidth]{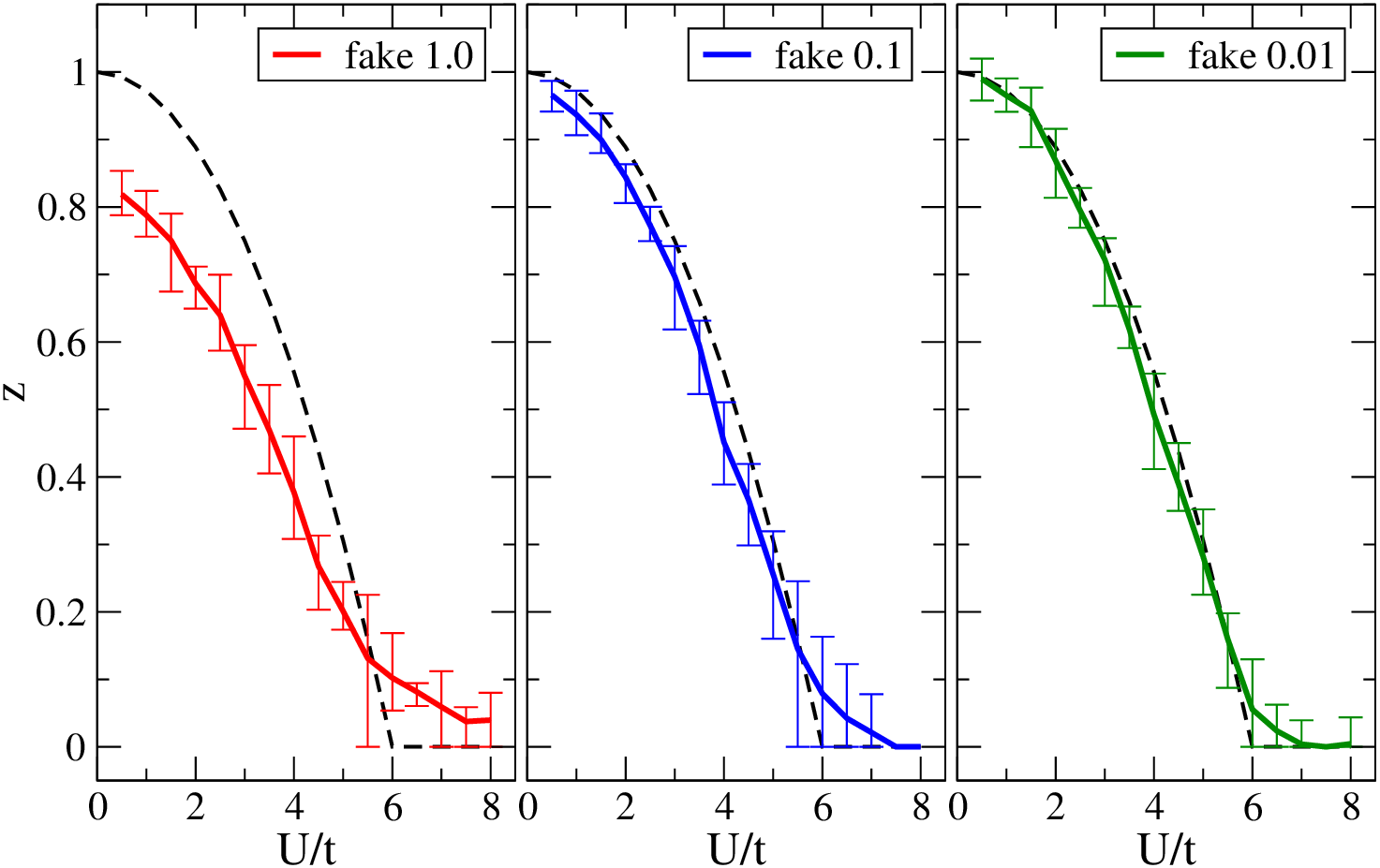}
\caption{Quasi-particle weight obtained from full self-consistent two-site DMFT simulations with fake backends based on the error rates and coherence times of IBMQ Ehningen without scaling (fake 1.0), with scaling by a factor of 0.1 (fake 0.1) and a factor of 0.01 (fake 0.01). The solid lines mark the arithmetic mean values of ten runs, and the bars indicate the maximum and minimum of the obtained values. The exact result is shown as black dashed line. 
}\label{Fig:FakeBackend}
\end{figure}

For the next step towards an implementation on real QC hardware, we performed simulations of the phase diagram with a fake version of the IBMQ Ehningen \cite{ErrorValues} using \(10\,k\) shots.
Using the fake backend with the full noise model we reproduce the behavior in the weak coupling regime, but fail to reproduce the transition to the insulating phase, cf.\ left panel of Fig.\ \ref{Fig:FakeBackend}. In order to investigate the influence of noise on the results, we created fake backends with scaled noise. By a scale factor of 0.1 and 0.01 we refer to a scaling of gate and SPAM noise by a factor of 0.1 and 0.01 and an elongation of the coherence times \(T_1\) and \(T_2\) by a factor of 10 and 100, respectively. Performing the simulation with a noise factor of 0.1, the phase diagram is well captured, although the \(z\)-values are systematically underestimated. This is most probably due to the systematic overestimation of the eigenenergies, which shift the peaks of the self-energy closer to \(\omega=0\). Finally, by scaling the errors by a factor of 0.01, the phase diagram can be reproduced very well. However, as in the previous cases, our approach does not give the sharp Mott transition. Note that even in this case statistics are required and the good agreement of the displayed curve with the exact result is obtained from averaging individual data points with a spread as indicated by the error bars.


\subsection{Discussion}\label{Sec:Discussion}
We focused our investigations on the two-site DMFT model at half-filling with the aim of comparability with other studies. Our approach for the calculation of the GF as solution of the AIM and the fitting procedure for the determination of the quasi-particle weight can be generalized to doped two-site DMFT system as well as to larger systems. 

Considering the calculation of the GF, the particle number of the GS is initially not known for the doped two-site DMFT. Therefore, an ansatz covering the Hilbert space of arbitrary particle number is required for the corresponding VQE calculation, e.g.\ a hardware efficient ansatz. If the resulting particle number \(n=2\), we proceed as described above. However, if \(n=1\) or \(n=3\), the challenge is to determine the four \(n=2\) states with \(S_z=0\), as it requires an algorithm for excited states. The variational quantum deflation (VQD) \cite{Higgott2019} determines the \(m\)-th state by penalizing the overlap with all previously found \(m-1\) states. The subspace search VQE (SSVQE) \cite{Nakanishi2019} takes a set of orthogonal input states and optimizes one parameterized ansatz for all of them simultaneously, which leads to excited states as the orthogonality is preserved by unitary operators. The GS ansatz employed in our calculations can directly be used in both of these approaches. Alternatively matrix-based approaches like quantum subspace expansion (QSE) \cite{McClean2017} or quantum equation of motion (QEOM) \cite{Ollitrault2020,Rizzo2022} could be used, where the GS is expanded in an excitation basis and the resulting generalized eigenvalue problem is solved classically. \\

This approach also holds for larger systems with \(N\) orbitals. First, the GS and its particle number \(n\) is obtained from a VQE calculation with a particle number unrestricted ansatz. Then, the \(N_\textrm{e}=\binom{n + 1}{N}\) electron and \(N_\textrm{h}=\binom{n - 1}{N}\) hole states can be determined by VQD or SSVQE with a particle number preserving ansatz like the one used in our investigations. Based on those results, the GF can then be calculated by Eq.\ \eqref{Eq:ImpLehmann}.
However, the binomial growth in the number of required states poses a challenge when it comes to large system sizes. In the Lehmann representation approaches all these states need to be calculated explicitly. This is particularly challenging for variational approaches like the one presented here, as the costs of quantum computations grow significantly with the number of states. However, these approaches could be modified to select only the most relevant states for the GF. On the other hand, quantum subspace approaches like Quantum Subspace Expansion and Quantum Equation of Motion circumvent the growth in optimization costs. However, the size of the matrices to be sampled still grows in the order of \(N_\textrm{e/h}^2\) with the number of required states. An approach in terms of the Krylov basis \cite{Jamet2021, Jamet2022} shows a beneficial scaling behavior, as the iteratively calculated coefficients appear to decrease independently of the system size and provide a natural termination criterion. In comparison, the scaling based on the real-time GF depends only on the number of time steps and not on the system size. However, longer times and probably smaller time steps need to be simulated for larger systems in order to improve the resolution of the Fourier transformation. Therefore, quantum circuits of a length beyond the capabilities of current NISQ hardware would be required, thus leaving these approaches for future better hardware or error-corrected QC. 

For the determination of the self-energy and the quasi-particle weight a good quality of the self-energy is essential. For time evolution approaches, this is achieved by a fit of the noisy quantum computer results to the analytically known form of the GF of this model (see \cite{Jaderberg2020, Keen2020}). For approaches based on the Lehmann representation we showed in Sec.\ \ref{Sec:GF} that the pure shot noise leads to unphysical results even for \(100 k\) shots. Thus, a regularization procedure for the measured values was suggested and applied by Rungger et al\cite{Rungger2019}. They showed an anlytical integration of this regularization for the special case of the two-site DMFT at half-filling and pointed towards a general formulation with classical optimization techniques. However, a test of their approach yielded regularization-corrections to expectation values and transition rates, which are of comparable size to the respective measured quantity itself or even larger. Therefore, a confidence in the final outcome of this correction procedure is highly questionable.\\

In this paper we suggested an alternative, by determining physical features and artifacts in the self-energy and by fitting a generic function to it. As the comparison of Fig.\ \ref{Fig:QC_results} and Tab.\ \ref{Tab:QC_z} shows, our \(\tan\)-fit approach works for peaks which are not symmetric with respect to \(\omega = 0\) due to proper scaling and shift of the fitting interval. However, if the peaks themselves significantly differ in their shape, the fitting function requires an appropriate adaptation to account for this shape asymmetry.


\section{Conclusion}\label{Sec:Outlook}

In this work we explored the practical power of an available quantum computer to obtain the Green's function of a fermionic quantum system through its Lehmann representation. The general procedure consists of (i) determining the relevant eigenstates by a quantum-classical hybrid variational algorithm and (ii) measuring the transition rates by employing a quantum computer. In this general form, it can be applied to quantum systems of arbitrary size and filling. Here, we explicitly considered the two-site DMFT model at half-filling and used the VQE approach to determine the relevant eigenstates. Thus, knowledge of the electron number and spin of the target states was only exploited for the variational state determination but not in the construction of the GF or the calculation of the quasi-particle weight. We showed that a self-consistent hybrid quantum-classical solution of the DMFT model is in principle possible. By calculations (i) with only stochastic noise, (ii) with simulated device noise, and (iii) on a real IBMQ machine we extracted three present obstacles for full self-consistent DMFT calculations on state-of-the-art quantum computers based on the Lehmann-representation approach and a state determination by variational approaches: 

(1) The limited accuracy of the expectation values of the eigenenergies and transition rates, already resulting from statistical noise, leads to an unphysical two-peak structure in the self-energy on the real frequency axis, explicitly around \(\omega=0\). As this prohibits the determination of the quasi-particle weight from the derivative, we introduced a fitting approach to cure the artificial two-peak structure enclosing \(\omega=0\). By self-consistent calculations we showed that this approach is able to reproduce the correct phase diagram of the model including the Mott transition, even in the presence of shot-noise with at least \(10\,k\) shots. We note that there are alternative approaches for the determination of the quasi-particle weight, like calculations in Matsubara frequencies or using the Kramers-Kronig relation to convert it into an integral over the imaginary part of the self-energy \cite{Keen2020}.

(2) The device noise, like SPAM, dephasing and CNOT errors, leads to even less accurate results. While the fitting procedure introduced above can overcome the more pronounced two-peak structure around \(\omega=0\), its results are always too low, as even the physical peaks of the self-energy are shifted. Therefore, by performing self-consistent simulations with scaled noise of a real backend (with SPAM, dephasing and CNOT errors), we demonstrate that the noise rates of current quantum computers have to be improved by an order of magnitude in order to obtain numerical results with acceptable accuracy for the VQE-based determination of the GF in the Lehmann representation. However, with a post-processing IC-ZNE only for the GS the GF can be significantly improved, even on current NISQ devices which, however, requires additional QC resources.

(3) A self-consistent calculation on real quantum hardware based on the Lehmann representation requires a large amount of quantum resources, when all states are determined by variational algorithms. For the two-site DMFT model these are 9 VQE simulations and the calculation of 8 transition matrix elements for the calculation of one GF. Without any parallelization this requires several hours on current QC hardware. As this is necessary in each of the at least 10 iterations of the DMFT self-consistency loop, it would run on a single QPU for several days. 
Although possible in principle, the financial costs due to the current pricing for QPU time renders such a calculation hardly feasible in practice. 
As pointed out in Sec.\ \ref{Sec:Discussion}, a large number of states has to be calculated for larger systems. The VQD \cite{Higgott2019} has a polynomial scaling and can not be parallelized, due to its iterative structure. Contrary, the SSVQE \cite{Nakanishi2019} scales with the number of states required and is parallelizable. The semi-classical matrix-based approaches like QSE \cite{McClean2017} or QEOM \cite{Ollitrault2020,Rizzo2022} may show a better scaling behavior, as they require only one VQE optimization of the GS and the measurement of transition matrix elements. However, the growth in the subspace leads to a large number of Pauli-terms that have to be calculated.

These points (1)--(3) suggest, that solving a general DMFT model with satisfactory accuracy within a reasonable amount of computation time on a gate-based QC via the Lehmann approach remains a challenge for the near future. 

In conclusion, hybrid quantum-classical approaches for the impurity solver in DMFT simulations are promising candidates to improve the results in terms of accuracy and speed. However, further development on the scalability to larger systems and noise-resilience are required for a general applicability. 

\section{Data availability}
The data cannot be made publicly available upon publication because they are not available in a format that is sufficiently accessible or reusable by other researchers. The data that support the findings of this study are available upon reasonable request from the authors.

\section{Acknowledgments}
This work has been funded by the Ministry of Economic Affairs, Labour and Tourism Baden Württemberg through the Competence Center Quantum Computing Baden-Württemberg (KQCBW), projects QuESt and QuESt+.

\bibliography{two_site_paper_lit} 

\end{document}